\documentclass[10pt]{iopart}
\pdfoutput=1

\expandafter\let\csname equation*\endcsname\relax
\expandafter\let\csname endequation*\endcsname\relax
\usepackage{amsmath}

\usepackage{hyperref} 

\usepackage[english]{babel}
\usepackage{color}
\usepackage{epsfig}
\usepackage{graphicx}
\usepackage{bm}
\usepackage{times}
\usepackage{amsfonts}
\usepackage{amssymb}
\usepackage{color}
\usepackage{leftidx}

\newcommand{\media}[1]{\left\langle #1 \right\rangle}
\newcommand{\bra}[1]{\langle #1 |} \newcommand{\ket}[1]{| #1 \rangle}
\newcommand{\scp}[2]{\langle #1 | #2 \rangle}
\newcommand{\braket}[3]{\langle #1 | #2 | #3 \rangle}

\def\RE{\mathop{\rm Re}} %
\newcommand{\Span}{\mathrm{span}} %

\newcommand{\rev}[1]{{\color{black}#1}}

\begin{document}

\title[First-order quantum phase transitions as condensations in the
space of states]{First-order quantum phase transitions as
  condensations in the space of states}

\author{Massimo Ostilli} \address{Instituto de F\'isica, Universidade
  Federal da Bahia, Salvador, Brazil}

\author{Carlo Presilla} \address{Dipartimento di Fisica, Sapienza
  Universit\`a di Roma, Piazzale A. Moro 2, Roma 00185, Italy}
\address{Istituto Nazionale di Fisica Nucleare, Sezione di Roma 1,
  Roma 00185, Italy}

\vspace{10pt}
\begin{indented}
\item[]\today
\end{indented}

\begin{abstract}
  We demonstrate that a large class of first-order quantum phase
  transitions, namely, transitions in which the ground state energy
  per particle is continuous but its first order derivative has a jump
  discontinuity, can be described as a condensation in the space of
  states.  Given a system having Hamiltonian $H=K+gV$, where $K$ and
  $V$ are \rev{two non commuting} 
  operators
  acting on the space of
  states $\mathbb{F}$, we may always write
  $\mathbb{F}=\mathbb{F}_\mathrm{cond} \oplus
  \mathbb{F}_\mathrm{norm}$ where $\mathbb{F}_\mathrm{cond}$ is the
  subspace spanned by the eigenstates of $V$ with minimal eigenvalue
  and $\mathbb{F}_\mathrm{norm}=\mathbb{F}_\mathrm{cond}^\perp$.  If,
  in the thermodynamic limit, $M_\mathrm{cond}/M \to 0$, where $M$ and
  $M_\mathrm{cond}$ are, respectively, the dimensions of $\mathbb{F}$
  and $\mathbb{F}_\mathrm{cond}$, the above decomposition of
  $\mathbb{F}$ becomes effective, in the sense that the ground state
  energy per particle of the system, $\epsilon$, coincides with the
  smaller between $\epsilon_\mathrm{cond}$ and
  $\epsilon_\mathrm{norm}$, the ground state energies per particle of
  the system restricted to the subspaces $\mathbb{F}_\mathrm{cond}$
  and $\mathbb{F}_\mathrm{norm}$, respectively:
  $\epsilon=\min\{\epsilon_\mathrm{cond},\epsilon_\mathrm{norm}\}$.
  It may then happen that, as a function of the parameter $g$, the
  energies $\epsilon_\mathrm{cond}$ and $\epsilon_\mathrm{norm}$ cross
  at $g=g_\mathrm{c}$. In this case, a first-order quantum phase
  transition takes place between a condensed phase (system restricted
  to the small subspace $\mathbb{F}_\mathrm{cond}$) and a normal phase
  (system spread over the large subspace $\mathbb{F}_\mathrm{norm}$).
  Since, in the thermodynamic limit, $M_\mathrm{cond}/M \to 0$, the
  confinement into $\mathbb{F}_\mathrm{cond}$ is actually a
  condensation in which the system falls into a ground state
  orthogonal to that of the normal phase, something reminiscent of
  Anderson's orthogonality catastrophe (P.~W.~Anderson,
  Phys. Rev. Lett. \textbf{18}, 1049 (1967)).  The outlined mechanism
  is tested on a variety of benchmark lattice models, including spin
  systems, free fermions with non uniform fields, interacting fermions
  and interacting hard-core bosons.
\end{abstract}

\section{Introduction}
Unlike classical phase transitions, which are based on a competition
between entropy maximization and energy minimization, tuned by varying
the temperature, quantum phase transitions (QPT) are characterized by
a competition between two qualitatively different ground states (GSs)
reachable by varying the Hamiltonian parameters at zero
temperature~\cite{Sondhi,Sachdev,Vojta,Dutta}.  Typically, one has to
compare the effects of two non commuting operators.  To be specific,
let us consider a lattice model with $N$ sites and $N_\mathrm{p}$
particles described by a Hamiltonian
\begin{align}
  \label{H}
  H=K+g V,
\end{align}
where $K$ and $V$ are Hermitian non commuting operators, and $g$ a
free dimensionless parameter.  One can represent $H$ in the eigenbasis
of $V$.  In such a case, it is natural to call $V$ a \rev{``potential''}
operator, and $K$ a \rev{``hopping''} operator. Let us suppose that both $K$ and
$V$ scale linearly with the number of particles $N_\mathrm{p}$.  Since
in the two opposite limits $g\to 0$ and $g\to \infty$, the GS of the
system tends to the GS of $K$ and $V$, respectively, one wonders if,
in the thermodynamic limit, a QPT takes place at some intermediate
critical value of $g$: $g_\mathrm{c}=O(1)$.  In fact, an argument
based on the ``avoided-crossing-levels''~\cite{Landau} suggests that a
possible abrupt bending of the GS energy of $H$ occurs.  However,
there is no exact way to apply this scheme and, by varying $g$, three
possibilities remain open: i) there is no QPT; ii) there exists a
$g_\mathrm{c}$ where a second-order transition takes place; iii) there
exists a $g_\mathrm{c}$ where the first derivative of the GS energy
makes a finite jump.  Let us discuss briefly these scenarios.

i) Here we mention only that, in principle, there could be no QPT at
all, or even a QPT with no singularity~\cite{Dutta}.

ii) Within some extent, Landau's theory of classical critical
phenomena offers a universal approach also to second-order QPTs via
the quantum-classical mapping, according to which the original quantum
model in $d$ dimensions is replaced by an effective classical system
in $d+z$ dimensions~\cite{Suzuki,Sondhi,Dutta}, $z$ being the
dynamical critical exponent.  Hence, for second-order QPTs, concepts
and tools originally defined for classical critical phenomena find a
quantum counterpart and the main issue concerns the competition
between classical and quantum fluctuations.

iii) A different situation occurs for first-order QPTs for which
\rev{a universal mechanism seems lacking}~\cite{Pfleiderer}.  As for
the classical case, first-order QPTs can result from \rev{the finite
  jump of the order parameter when crossing the coexistence line of
  two different phases} that originate from the same critical point of
a second-order transition~\cite{Continentino}. Notice that, for such a
scenario to occur at zero temperature, one needs that $H$ (or the
corresponding Lagrangian) depends on at least two independent
parameters (say $g_1$ and $g_2$).  \rev{ In these first-order
  transitions, universality reflects on finite size
  scaling~\cite{Pelissetto}. Some kind of universality of first-order
  QPTs is expected also in systems of vector spin models with a
  sufficiently large number of components, as in the 1D quantum Potts
  model~\cite{PelissettoPotts}.  In general, however, for genuine
  first-order QPTs driven by a single parameter $g$, i.e., those that
  do not originate from the critical point of a second-order transition
  (as, e.g., in the case of frustrated~\cite{Bose}, mean-field, and
  random spin systems~\cite{Vidal,OP,Krzakala,Krzakala1}), or those
  for which there is no evident classical analog, it is not clear
  which universal mechanism, if any, is at their basis.  Lack of
  universality has been in fact observed in the gap $\Delta$ (the
  difference between the two lowest energy levels) of certain
  systems~\cite{Nishimori}.  In some cases $\Delta$ does not take the
  absolute minimum at $g_\mathrm{c}$ (see~\ref{Spec2}). There are
  even systems where $\Delta$ remains finite, as in certain
  topological second-order~\cite{Ezawa,Rachel} and
  first-order~\cite{Amaricci,Roy} QPTs.}

In this paper, we test a theory concerning a large class of
first-order QPTs that lead to many-body condensation thorough a
counter-intuitive mechanism having no classical analog \rev{and that
  can be interpreted as a many-body Anderson's orthogonality
  catastrophe~\cite{Anderson}. The approach provides also an efficient
  criterion for localizing the critical point}.  We first formulate
the theory in general terms, regardless of the details of $K$ and $V$
\rev{(where we allow for the presence of also more than one parameter)},
\rev{then} we test it on several specific models: spin systems, free
fermions in a heterogeneous external field, interacting fermions and
interacting hard-core bosons, with both open and periodic boundary
conditions.  \rev{A comparison with the fidelity
  method~\cite{Fidelity_Gu} is also discussed.}  The paper is equipped
with Appendices that illustrate some immediate extensions of the
theory and provide further critical checks.  Here, we discuss our
theory at zero temperature while its finite temperature counterpart
will be reported elsewhere.
 
\section{Main result}
Consider a system with Hamiltonian as in Eq.~(\ref{H}), and let
$\{ \ket{n} \}$ be a complete orthonormal set of eigenstates of $V$:
$V \ket{n} =V_n \ket{n}$, $n=1,\dots,M$.  We assume ordered potential
values $V_1 \leq V_2 \leq \dots \leq V_M$.  Let $M_\mathrm{cond}$ be
the degeneracy of the smallest potential
$V_\mathrm{min}=V_1=V_2= \dots =V_{M_\mathrm{cond}}$.  The Hilbert
space of the system, $\mathbb{F} = \Span \{ \ket{n} \}_{n=1}^{M}$,
equipped with standard complex scalar product $\scp{u}{v}$, can be
decomposed as the direct sum
$\mathbb{F}=\mathbb{F}_\mathrm{cond} \oplus \mathbb{F}_\mathrm{norm}$,
where
$\mathbb{F}_\mathrm{cond} = \Span \{ \ket{n}
\}_{n=1}^{M_\mathrm{cond}}$ and
$\mathbb{F}_\mathrm{norm} = \Span \{ \ket{n}
\}_{n=M_\mathrm{cond}+1}^{M} = \mathbb{F}_\mathrm{cond}^\perp$. Any
vector $\ket{u}\in\mathbb{F}$ can be uniquely written as
$\ket{u} = \ket{u_\mathrm{cond}} + \ket{u_\mathrm{norm}}$, where
$\ket{u_\mathrm{cond}} \in \mathbb{F}_\mathrm{cond}$ and
$\ket{u_\mathrm{norm}} \in \mathbb{F}_\mathrm{cond}^{\perp}$.
Finally, we define
\begin{align*}
  & E=\inf_{\ket{u}\in\mathbb{F}} \frac{\braket{u}{H}{u}}{\scp{u}{u}}
  \\
  & E_\mathrm{cond}=\inf_{\ket{u}\in\mathbb{F}_\mathrm{cond}}
    \frac{\braket{u}{H}{u}}{\scp{u}{u}},
  \\
  & E_\mathrm{norm}=\inf_{\ket{u}\in\mathbb{F}_\mathrm{norm}}
    \frac{\braket{u}{H}{u}}{\scp{u}{u}}.
\end{align*}
Clearly, $E$ is the GS energy of the system and by construction
$E\leq \min\{E_\mathrm{cond},E_\mathrm{norm}\}$.  Less trivial is to
understand the relation among $E$, $E_\mathrm{cond}$ and
$E_\mathrm{norm}$ in the thermodynamic limit.

To properly analyze this limit, let us consider systems consisting of
$N_\mathrm{p}$ particles in a lattice with $N$ sites and assume that
the lowest eigenvalues of $K$ and $V$ scale linearly with
$N_\mathrm{p}$, at least for $N_\mathrm{p}$ large.  The thermodynamic
limit is defined as the limit $N,N_\mathrm{p}\to\infty$ with
$N_\mathrm{p}/N=\varrho$ constant. Because of the assumed scaling
properties, the energies $E(N,N_\mathrm{p})$,
$E_\mathrm{cond}(N,N_\mathrm{p})$ and
$E_\mathrm{norm}(N,N_\mathrm{p})$ diverge linearly with the number of
particles, therefore, if divided by $N_\mathrm{p}$, they have finite
thermodynamic limits which depend on the chosen density $\varrho$. We
call these limits $\epsilon(\varrho)$,
$\epsilon_\mathrm{cond}(\varrho)$, and
$\epsilon_\mathrm{norm}(\varrho)$~\cite{Note1}.

We state that, under the above scaling conditions on $K$ and $V$,
\begin{align}
  \label{QPT}
  \mbox{if } \lim_{N,N_\mathrm{p}\to\infty,N_\mathrm{p}/N=\varrho}
  {M_\mathrm{cond}/M}=0, \qquad
  \mbox{then } \epsilon(\varrho)=
  \min\{\epsilon_\mathrm{cond}(\varrho),\epsilon_\mathrm{norm}(\varrho)\}.
\end{align}
Equation (\ref{QPT}) establishes the possibility of a QPT between a
\textit{normal phase} characterized by the energy per particle
$\epsilon_\mathrm{norm}$, obtained by removing from $\mathbb{F}$ the
infinitely smaller sub-space $\mathbb{F}_\mathrm{cond}$, and a
\textit{condensed phase} characterized by the energy per particle
$\epsilon_\mathrm{cond}$ obtained by restricting the action of $H$
onto $\mathbb{F}_\mathrm{cond}$.  Note that the Hilbert space
dimension $M(N,N_\mathrm{p})$ diverges, generally in an exponential
way, with $N$ and $N_\mathrm{p}$.  The dimension $M_\mathrm{cond}$ may
or may not be a growing function of $N$ and $N_\mathrm{p}$.  In any
case, if $M_\mathrm{cond}/M\to 0$, in the space of the Hamiltonian
parameters the equation
\begin{align}
  \label{QCP}
  \epsilon_\mathrm{norm}(\varrho)= \epsilon_\mathrm{cond}(\varrho),
\end{align}
provides the coexistence surface of two phases, crossing which a QPT
takes place. If $H$ depends on a single parameter $g$, as in
Eq.~(\ref{H}), the coexistence surface reduces to a critical point
$g_\mathrm{c}$, which is given as the minimal solution, if any, of
Eq.~(\ref{QCP}) (in general, $\epsilon_\mathrm{norm}$ and
$\epsilon_\mathrm{cond}$ can be equal also for $g\neq g_\mathrm{c}$).
\rev{In virtue of Eq.~(\ref{QCP}), the transition is
first-order. In fact, $\epsilon_\mathrm{norm}$ and
  $\epsilon_\mathrm{cond}$ correspond to the lowest
  eigenvalues of two different matrix Hamiltonians so that,
  as functions of the Hamiltonian
  parameters, they are different, except possibly for a finite number of
  crossing points solution of Eq.~(\ref{QCP}).  Assuming that
  $\epsilon_\mathrm{norm}$ and
  $\epsilon_\mathrm{cond}$ are both everywhere analytic,
  implies that their first derivatives around any crossing point are, in general,
  also different. This is quite clear when there is only one
  Hamiltonian parameter: given two analytic functions of one variable,
  either they coincide everywhere, or their first derivatives at the
  crossing points (if any) are different. According to Ehrenfest
  classification this means that the QPT is of first-order.  }

\rev{Equations~(\ref{QPT}-\ref{QCP}) are quite general.
Whereas the paradigm of second-order QPTs
looks for changes of symmetries of the GS, the paradigm of the first-order QPTs determined 
by Eqs.~(\ref{QPT}-\ref{QCP}) looks for 
condensations of the GS into $\mathbb{F}_\mathrm{cond}$ (which may be accompanied by a broken symmetry or not).
The word ``condensation'' here refers to the fact that, in the condensed phase,
thanks to the condition $M_\mathrm{cond}/M\to 0$,
the GS occupies an infinitesimal portion of the full space of states. Depending on the structure of the space
$\mathbb{F}_\mathrm{cond}$, this abstract condensation in the Hilbert space can have different physical manifestations. 
As we shall show by specific examples,
in certain cases the condensation can realize through a partial or total freezing of the particles, and
even as an actual localization of matter.  
}
Such condensations were
first demonstrated in~\cite{OP} for two classes of models, the
uniformly fully connected models and the random potential systems.
For general systems, a formal proof based on the concept of sojourn
times in the subspaces $\mathbb{F}_\mathrm{cond}$ and
$\mathbb{F}_\mathrm{norm}$ is given in~\cite{qpt_dilution}. Here, we
provide a simple algebraic argument which goes as follows.  We start
with the obvious inequality
$\epsilon \leq \min
\{\epsilon_\mathrm{cond},\epsilon_\mathrm{norm}\}$, and demonstrate
that the opposite inequality holds too.  Let us evaluate $\epsilon$ as
the thermodynamic limit of
\begin{align}
  \label{inf_x1}
  &\frac{1}{N_\mathrm{p}} \inf_{\ket{u} \in \mathbb{F}}
    \frac{\braket{u}{H}{u}}{\scp{u}{u}}
    \nonumber\\
  &\qquad =    
    \frac{1}{N_\mathrm{p}} \inf_{0 \leq x \leq 1}
    \inf_{\ket{u_\mathrm{cond}} \in \mathbb{F}_\mathrm{cond}}
    \inf_{\ket{u_\mathrm{norm}} \in \mathbb{F}_\mathrm{norm}}
    \biggl(
    \frac{\braket{u_\mathrm{cond}}{H}{u_\mathrm{cond}}}
    {\scp{u_\mathrm{cond}}{u_\mathrm{cond}}} x
    +\frac{\braket{u_\mathrm{norm}}{H}{u_\mathrm{norm}}}
    {\scp{u_\mathrm{norm}}{u_\mathrm{norm}}} (1-x)
    \nonumber \\
  &\qquad\qquad +  \frac{
    \RE\braket{u_\mathrm{cond}}{K}{u_\mathrm{norm}}}
    {\sqrt{\scp{u_\mathrm{cond}}{u_\mathrm{cond}}
    \scp{u_\mathrm{norm}}{u_\mathrm{norm}}}} 2\sqrt{x(1-x)}
    \biggr),
\end{align}
where $\ket{u}=\ket{u_\mathrm{cond}}+\ket{u_\mathrm{norm}}$ with
$\scp{u_\mathrm{cond}}{u_\mathrm{norm}}=0$ and
$x=\scp{u_\mathrm{cond}}{u_\mathrm{cond}}/\scp{u}{u}$.  We find
\begin{align}
  \label{inf_x2}
  \epsilon \geq \inf_{0 \leq x \leq 1} \left( \epsilon_\mathrm{cond} x
  + \epsilon_\mathrm{norm} (1-x) + \beta 2\sqrt{x(1-x)} \right),
\end{align}
where $\beta$ is the thermodynamic limit of $B/N_\mathrm{p}$ and
\begin{align}
  B=\inf_{\ket{u_\mathrm{cond}} \in \mathbb{F}_\mathrm{cond}}
  \inf_{\ket{u_\mathrm{norm}} \in \mathbb{F}_\mathrm{norm}} \allowbreak
  \RE\braket{u_\mathrm{cond}}{K}{u_\mathrm{norm}}/ 
  \sqrt{\scp{u_\mathrm{cond}}{u_\mathrm{cond}}
  \scp{u_\mathrm{norm}}{u_\mathrm{norm}}}.
\end{align}
Let
$\ket{\tilde{u}_\mathrm{cond}}=\sum_{n=1}^{M_\mathrm{cond}} c_n
\ket{n}$ and
$\ket{\tilde{u}_\mathrm{norm}}=\sum_{n=M_\mathrm{cond}+1}^{M} d_n
\ket{n}$ be the states of $\mathbb{F}_\mathrm{cond}$ and
$\mathbb{F}_\mathrm{norm}$ which realize this double infimum.  We now
prove that $\beta=0$.  Suppose, for simplicity, that $K$ is the sum of
$N_\mathrm{p}$ single-particle jump operators, i.e.,
$\braket{n}{K}{m}=-1$ if $m$ is one of the $N_\mathrm{p}$
configurational states first neighbor to $n$, and zero otherwise.  We
estimate
\begin{align*}
  \frac{|B|}{N_\mathrm{p}}
  \leq \frac{1}{N_\mathrm{p}}
  \sum_{n=1}^{M_\mathrm{cond}} \sum_{m=M_\mathrm{cond}+1}^{M}
  \left|c_n\right| \left|d_m\right| \left|\braket{n}{K}{m}\right| \sim
  \sqrt\frac{M_\mathrm{cond}}{M}
\end{align*}
provided that, as we expect normalizing the states
$\ket{\tilde{u}_\mathrm{cond}}$ and $\ket{\tilde{u}_\mathrm{norm}}$ to
1, $\left|c_n\right|\sim 1/\sqrt{M_\mathrm{cond}}$ and
$\left|d_m\right| \sim 1/\sqrt{M-M_\mathrm{cond}}$.  If, in the
thermodynamic limit, $M_\mathrm{cond}/M\to 0$, it follows that
$\beta=0$ and Eq.~(\ref{inf_x2}) gives
$\epsilon\geq\min\{\epsilon_\mathrm{cond},\epsilon_\mathrm{norm}\}$.
 
When $\min\{E_\mathrm{cond},E_\mathrm{norm}\}$ becomes, for
$N,N_\mathrm{p}$ finite but increasing, closer and closer to $E$,
$\max\{E_\mathrm{cond},E_\mathrm{norm}\}$ provides, although only
close to the critical point, a good approximation to $E'$, the energy
of the first excited state of $H$.  Whereas $E'$ is difficult to
evaluate numerically, $E_\mathrm{cond}$ and $E_\mathrm{norm}$, which
are both defined as GS energies of the system restricted to
$\mathbb{F}_\mathrm{cond}$ and $\mathbb{F}_\mathrm{norm}$, are a much
easier target, specially in Monte Carlo simulations (MCSs).  We
therefore define~\cite{Note1}
\begin{align}
  \label{Delta0}
  \Delta_0=\left|E_\mathrm{cond}-E_\mathrm{norm}\right|,
\end{align}
whose minimum allow us to locate in a simple way the critical point
when $N,N_\mathrm{p}$ are large enough.  However, in the cases in
which $\epsilon_\mathrm{norm}$ and $\epsilon$ overlap for all $g$ in
some interval (eventually infinite), rather than to cross just at
$g_\mathrm{c}$, it is convenient to locate $g_\mathrm{c}$ by analyzing
\begin{align}
  \label{Delta1}
  \Delta_1=|E_\mathrm{cond}-E|.
\end{align}
When possible, we compare $\Delta_0$ and $\Delta_1$ with
$\Delta=E'-E$, the ordinary gap. Notice that, in general, according to
Eq.~(\ref{QPT}), only $\Delta_0/N_\mathrm{p}$ and
$\Delta_1/N_\mathrm{p}$ vanish at $g=g_\mathrm{c}$. However, a plot of
$\Delta_0$ and $\Delta_1$ effectively allows for a precise
localization of $g_\mathrm{c}$.

In the following, we test Eqs.~(\ref{QPT}-\ref{QCP}) on several models
by means of numerical diagonalizations (NDs) and MCSs~\cite{MC}.  The
approach to the thermodynamic limit is studied by increasing the size
$N$ with $N_\mathrm{p}=\varrho N$ and $\varrho$ fixed.

\section{Grover Model}
Let us consider a set of $N$ spins with Hamiltonian
\begin{align}
  \label{GroverH}
  H=-\sum_{i=1}^N\sigma_i^x - g N \bigotimes_{i=1}^N
  \frac{1-\sigma_i^z}{2},
\end{align}
where $\sigma_i^x$ and $\sigma_i^z$ are the Pauli matrices acting on
the $i$-th spin. In this model, in which $N_\mathrm{p}=N$, we have
$M=2^N$ and
$\mathbb{F}=\Span \{|s_1\rangle \otimes \dots \otimes|s_N\rangle\}$,
where $|s_i\rangle$, with $s_i=\pm 1$, are the eigenstates of
$\sigma_i^z$.  Equation~(\ref{GroverH}) is of interest as a benchmark
model in quantum information theory, and corresponds to the quantum
version of the classical search problem~\cite{Grover,Bennet}, where a
single target state must be found over a set of $M$ unstructured
states. Notice that no efficient MCSs exist for this model, the form
of the potential being the worst case scenario for any hypothetical
importance sampling~\cite{Ceperley}.  The model is also of interest to
quantum adiabatic algorithms~\cite{QAD}.  In~\cite{OP} we solved a
random version of~(\ref{GroverH}), where the second term of $H$ is
built by randomly assigning the value $-gN$ to a single state of
$\mathbb{F}$ and a quenched average over many independent realizations
is taken at the end.  The present non-random model provides the
simplest paradigmatic example that illustrates the role and the
validity of Eqs. (\ref{QPT}-\ref{QCP}).

\begin{figure}[t]
  \centerline{ \includegraphics[width=\columnwidth, clip]{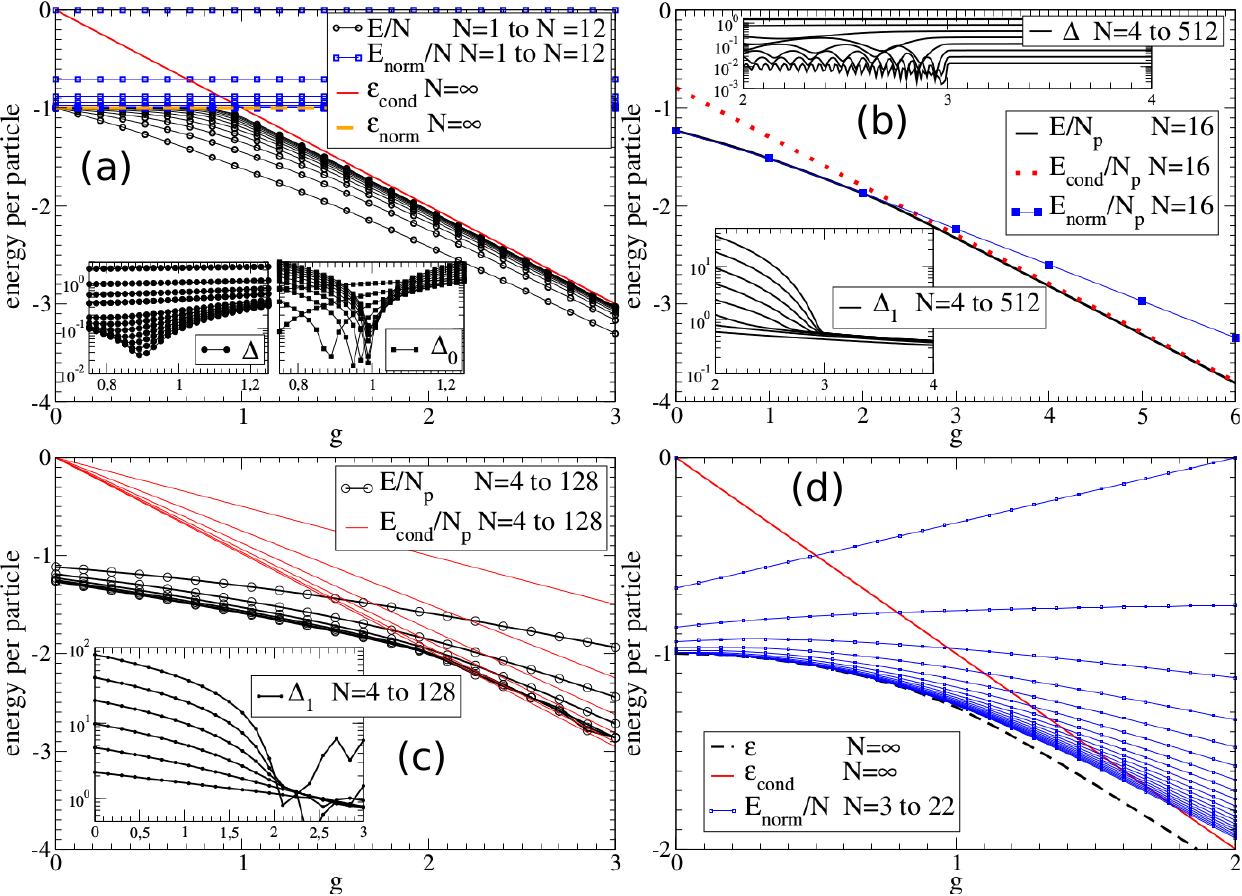}}
  \caption{ (\textbf{a}) GS energies per particle obtained from NDs,
    as a function of $g$, for the model described by
    Eq. (\ref{GroverH}).  Circles: $E/N$ for $N=1$ to $12$ (lowest to
    highest plot), lines are guides for the eyes. Squares:
    $E_\mathrm{norm}/N$ for $N=1$ to $12$ (highest to lowest).  The
    thermodynamic limits, Eqs. (\ref{GroverH1}), are represented by
    straight lines, $\epsilon_\mathrm{norm}$ (thick horizontal dashed
    line), and $\epsilon_\mathrm{cond}$ (solid line), crossing at the
    critical point $g_\mathrm{c}=1$.  Left Inset: gap $\Delta$ as a
    function of $g$, for $N=1$ to $12$ (highest to lowest). Right
    Inset: the function $\Delta_0$, Eq. (\ref{Delta0}), for $N=1$ to
    $12$ (about leftest to rightest).  (\textbf{b}) GS energies per
    particle for the model of Eq. (\ref{FermionH}) with
    $N_\mathrm{p}=N/2$ and $N_\mathrm{imp}=N/4$.  Upper Inset: gap
    $\Delta$, for $N=4,\dots,512$ via powers of 2 (highest to lowest).
    Lower Inset: the function $\Delta_1$, for $N=4,\dots,512$ via
    powers of 2 (highest to lowest).  The curves
    $E_\mathrm{cond}/N_\mathrm{p}$ are obtained from Eq. (\ref{eFsub})
    whereas $E_\mathrm{norm}/N_\mathrm{p}$ from MCSs. Here
    $g_\mathrm{c}\simeq 3$.  (\textbf{c}): GS energies per particle
    for the model of Eq. (\ref{FermionHA}) with $N_\mathrm{p}=N/2$ for
    $N=4,\dots,128$ via powers of 2 (highest to lowest).  Solid lines:
    $E_\mathrm{cond}/N_\mathrm{p}$ from Eq. (\ref{MFsubqA}).  Circles:
    $E/N_\mathrm{p}$ from MCSs.  Inset: $\Delta_1$ for $N=4,\dots,128$
    via powers of 2 (lowest to highest).  Here
    $g_\mathrm{c}\simeq 2.0$.  (\textbf{d}): GS energies per particle
    for the Ising model, Eq. (\ref{Ising}):
    $\epsilon_\mathrm{cond}=-g$ (continuous line), $\epsilon_{}$
    (dashed line), $E_\mathrm{norm}/N$ (line with points) for $N$= 3
    to 22 (top to bottom).}
  \label{fig1}
\end{figure}

Comparing Eq.~(\ref{GroverH}) with Eq.~(\ref{H}), we see that
$K=-\sum_i\sigma_i^x$ and $V=-N \otimes_{i=1}^N(1-\sigma_i^z)/2$.  The
potential $V$ has its minimal eigenvalue in correspondence with the
state
$\ket{1} \equiv \ket{s_1=-1} \otimes \dots \otimes \ket{s_N=-1}$,
namely, $V_\mathrm{min}=V_1=-N$, whereas $V_n=0$ for $n=2,\dots,M$.
We thus have $M_\mathrm{cond}=1$,
$\mathbb{F}_\mathrm{cond}=\Span\{\ket{1}\}$, $E_\mathrm{cond}=-gN$ and
$\epsilon_\mathrm{cond}=-g$.  Consider now the GS of $H$ in
$\mathbb{F}_\mathrm{norm}=\mathbb{F}_\mathrm{cond}^{\perp}$. For $N$
finite, we are not able to analytically calculate $E_\mathrm{norm}$.
However, we observe that, since
$\ket{1}\notin \mathbb{F}_\mathrm{norm}$, $E_\mathrm{norm}$ cannot
depend on $g$.  Hence, for $E_\mathrm{norm}$ there is no QPT and we
can apply Eq.~(\ref{QPT}) to obtain
$\epsilon_\mathrm{norm}=\lim_{N\to\infty} E(g=0)/N=-1$.  In
conclusion,
\begin{align}
  \label{GroverH1}
  M_\mathrm{cond}/M=2^{-N}, \qquad \epsilon_\mathrm{cond}=-g, \qquad
  \epsilon_\mathrm{norm}=-1,
\end{align}
and applying Eq.~(\ref{QPT}) we find
\begin{align}
  \label{GroverH2}
  \epsilon_{}=\left\{
  \begin{array}{l}
    -1, \qquad g<1, \\
    -g, \qquad g\geq 1.
  \end{array}
  \right.
\end{align}
Figure~\ref{fig1}(\textbf{a}) shows the results from NDs.  As $N$
grows, the GS energy per spin tends to the curve $\epsilon$, predicted
by Eq.~(\ref{GroverH2}), with a finite discontinuity in its first
derivative at $g_\mathrm{c}=1$.  Figure~\ref{fig1}(\textbf{a}) also
shows that, as $N$ increases, $\Delta(g)$ and $\Delta_0(g)$ take their
minima at $g$ closer and closer to $g_\mathrm{c}$.

\rev{
\begin{figure}[t]
  \centerline{ \includegraphics[width=0.6\columnwidth, clip]{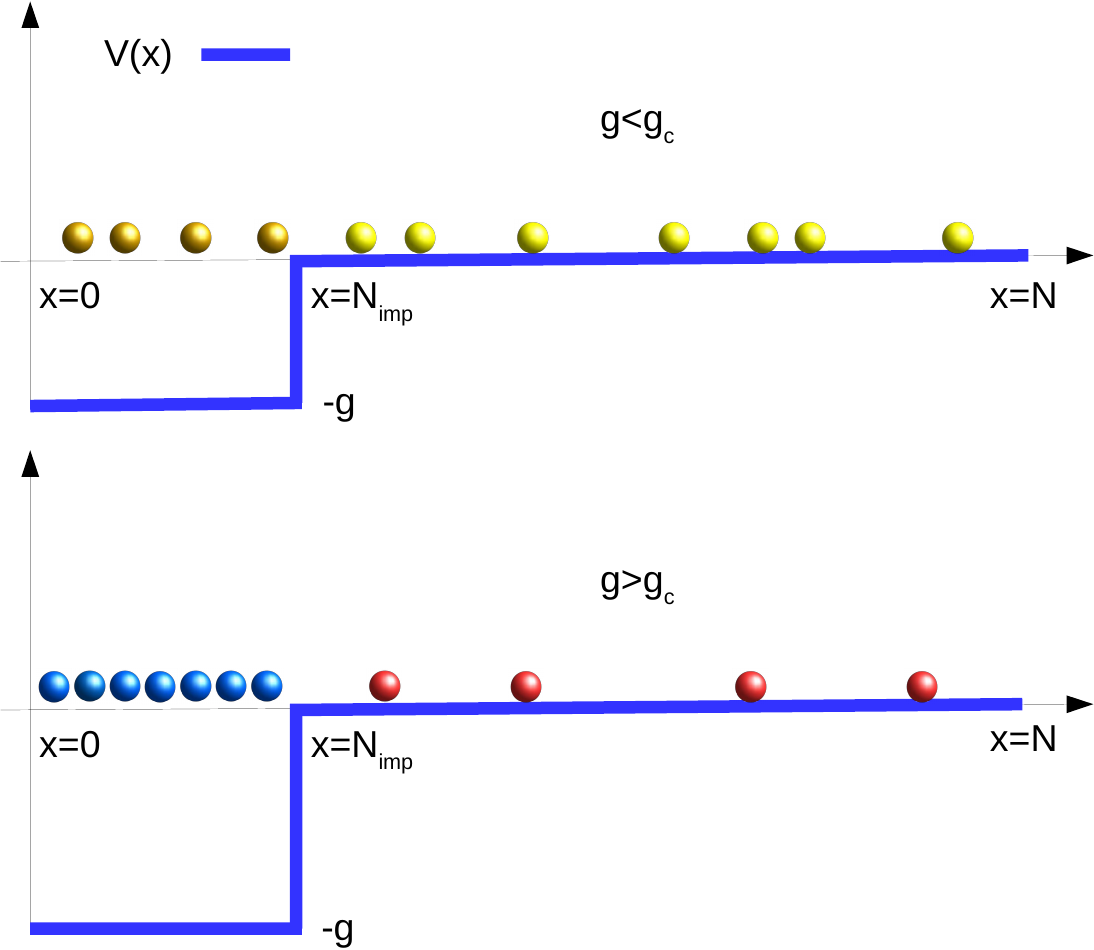}}
  \caption{ \rev{Pictorial view of the GS of the model given by
      Eq. (\ref{FermionH}) representing $N_\mathrm{p}$ free spinless
      electrons confined in a segment of length $N$ and in the
      presence of a non uniform external field. Here, the GS is the
      antisymmetrized product of $N_\mathrm{p}$ suitable
      single-particle states, each being eigenstate of a
      single-particle Hamiltonian with hopping operator $K$ and a step
      potential operator $V(x)$ characterized by a well of depth $g$
      and length $N_\mathrm{imp}$.  Top panel: For $g<g_c$ the GS is
      the antisymmetrized product of $N_\mathrm{p}$ quasi free
      electrons waves confined in the range $0\leq x\leq N$, with some
      slightly more localized than others in the range
      $0\leq x \leq N_\mathrm{imp}$ (sketched as light and dark yellow
      balls). Bottom panel: for $g>g_c$ the GS is the antisymmetrized
      product of $N_\mathrm{p}-N_\mathrm{imp}$ free electrons waves
      confined in the range $N_\mathrm{imp}<x\leq N$ (sketched as red
      balls) times the antisymmetrized product of $N_\mathrm{imp}$
      frozen position states over the range
      $0\leq x\leq N_\mathrm{imp}$ (sketched as blue balls).  Notice
      that using OBC or PBC does not affect the picture, see Appendix
      B for details.}}
  \label{fig_model_2}
\end{figure}
}

\section{Spinless fermions in 1D with a nonuniform external field}
\label{spinless.fermions.with.external.field}
Let us consider $N_\mathrm{p}$ spinless fermions in a 1D chain of
$N\geq N_\mathrm{p}$ sites with open boundary conditions (OBC). The
advantage of choosing OBC stems from the fact that, for fermions in
1D, there is no \textit{sign-problem} in MCSs~\cite{SignP}; in
\ref{OBCPBC} we discuss the case of periodic boundary conditions
(PBC). The Hamiltonian is
\begin{align}
  \label{FermionH}
  H=-\sum_{i=1}^{N-1}\left(c^\dag_ic_{i+1}+c^\dag_{i+1}c_i\right)
  -g\sum_{i=1}^{N_\mathrm{imp}}c^\dag_ic_i,
\end{align}
where $c_i$ are fermionic annihilation operators and
$N_\mathrm{imp}\leq N_\mathrm{p}$ is the number of
\textit{impurities}, or the number of sites where an external field
applies.  For simplicity, we choose to have these impurities in the
first $N_\mathrm{imp}$ sites of the chain.  This choice is not
restrictive but allows to calculate $\epsilon_\mathrm{cond}$ more
easily.  We consider the half-filling case $N_\mathrm{p}=N/2$ with $N$
even, so that $M=\binom{N}{N/2}$.  Since $H$ is quadratic in the
fermionic operators, the corresponding eigenvalue problem can be
exactly solved by diagonalizing the associated $N\times N$ Toeplix
matrix $\bm{A}$, whose non zero elements are $A_{i+1,i}=A_{i,i+1}=-1$,
for $i=1,\dots,N-1$, and $A_{i,i}=-g$, for $i=1,\dots,N_\mathrm{imp}$.
The eigenvalues of $\bm{A}$ are single particles energies, which,
summed up according to Pauli's principle, form the
$N_\mathrm{p}$-particle eigenvalues of $H$. The matrix $\bm{A}$ can be
numerically diagonalized for quite large sizes $N$ and we can evaluate
the exact gap as a further benchmark of the theory.

For $N_\mathrm{imp}=N_\mathrm{p}$, the minimal potential occurs in
correspondence with the single state
$|1\rangle=c^\dag_1c_1\dots
c^\dag_{N_\mathrm{p}}c_{N_\mathrm{p}}|\bm{0}\rangle$, where
$|\bm{0}\rangle$ is the vacuum state and
$V_\mathrm{min}=V_1=-N_\mathrm{p}$.  For
$N_\mathrm{imp}<N_\mathrm{p}$, instead, $V_\mathrm{min}$ is
degenerate, and $\mathbb{F}_\mathrm{cond}$ spans those states in which
$N_\mathrm{imp}$ fermions occupy the first $N_\mathrm{imp}$ sites. We
have
\begin{align}
  \label{MFsub}
  &M_\mathrm{cond}=\binom{N-N_\mathrm{imp}}{N_\mathrm{p}-N_\mathrm{imp}},  \\
  \label{eFsub}
  &\frac{E_\mathrm{cond}}{N_\mathrm{p}}=
    -g\frac{N_\mathrm{imp}}{N_\mathrm{p}} +\frac{E^{(0)}\left(
    {N-N_\mathrm{imp}},{N_\mathrm{p}-N_\mathrm{imp}}
    \right)}{N_\mathrm{p}},
\end{align}
where $E^{(0)}(N-N_\mathrm{imp},N_\mathrm{p}-N_\mathrm{imp})$ is the
GS energy of a system of $N_\mathrm{p}-N_\mathrm{imp}$ free spinless
fermions in a 1D lattice of $N-N_\mathrm{imp}$ sites with OBC, whose
single-particle energies are
$e^{(0)}_l=-2 \cos\left(\pi l/(N-N_\mathrm{imp}+1)\right)$, with
$l=1,\dots,N-N_\mathrm{imp}$.  In the normal phase the situation is
less simple, for $\mathbb{F}_\mathrm{norm}$ spans those states in
which no more than $N_\mathrm{imp}-1$ fermions occupy the impurity
sites.  This is equivalent to the action of a nonquadratic Hamiltonian
and we resort to MCSs to evaluate $E_\mathrm{norm}$.

Using Eq.~(\ref{MFsub}), it is easy to check that, in the
thermodynamic limit, $M_\mathrm{cond}/M\to 0$ for any non zero
fraction $N_\mathrm{imp}/N_\mathrm{p}$. In this case, we expect a
first-order QPT to take place if Eq.~(\ref{QCP}) has solution.  In
Fig.~\ref{fig1}(\textbf{b}) we report the analysis of the case
$N_\mathrm{imp}=N_\mathrm{p}/2$, while in \ref{OBCPBC} we show the
case $N_\mathrm{imp}=N_\mathrm{p}$. In both cases, Eq.~(\ref{QPT}) is
confirmed and a QPT takes place at the point $g_\mathrm{c}$ solution
of Eq.~(\ref{QCP}).  Interestingly, unlike the previous model, as $N$
increases, $\epsilon_\mathrm{norm}$ approaches $\epsilon$ in both the
normal and the condensed phases.  For visual convenience,
Fig.~\ref{fig1}(\textbf{b}) shows the behavior of $\epsilon$ and
$\epsilon_\mathrm{norm}$ only for one size value, the thermodynamic
limit being quickly approached in this model.  The plot of $\Delta_1$
(lower Inset) shows that the study of this quantity allows for an
excellent location of $g_\mathrm{c}$ in perfect agreement with the
analysis from the ordinary gap $\Delta$ (upper Inset).

\rev{This example is quite interesting: The Hamiltonian $H$, in
  Eq.~(\ref{FermionH}), does not contain any interaction among the
  particles and, in particular, there is no symmetry breaking,
  however, $H$ is still in the form~(\ref{H}), i.e., $H$ is the sum of
  two non commuting operators to which we can apply
  Eqs.~(\ref{QPT}-\ref{QCP}) and look for condensations. Remarkably,
  in this example the condensation corresponds to an actual
  localization of matter, as Fig. \ref{fig_model_2} shows.  }

\section{Spinless fermions in 1D with an attractive potential}
\label{spinless.fermions.with.attractive.potential}
Let us consider the following Hamiltonian of $N_\mathrm{p}$ fermions
in a 1D chain of $N\geq N_\mathrm{p}$ sites with OBC and an attractive
potential (as before, $g\geq 0$)
\begin{align}
  \label{FermionHA}
  H=-\sum_{i=1}^{N-1}\left(c^\dag_ic_{i+1}+c^\dag_{i+1}c_i\right)
  -g\sum_{i=1}^{N-1}c^\dag_ic_i c^\dag_{i+1}c_{i+1}.
\end{align}
Now, $V_\mathrm{min}$ corresponds to the closest packed configurations
of $N_\mathrm{p}$ fermions (one adjacent to the other one), and for
any finite value of $N_\mathrm{p}/N$, $M_\mathrm{cond}$ grows linearly
with $N$. Moreover, it is easy to see that $E_\mathrm{cond}$ has no
kinetic contributions. In conclusion,
\begin{align}
  \label{MFsubqA}
  M_\mathrm{cond}=N-N_\mathrm{p}, \qquad
  \frac{E_\mathrm{cond}}{N_\mathrm{p}} =
  -g\frac{N_\mathrm{p}-1}{N_\mathrm{p}}.
\end{align}
Fig.~\ref{fig1}(\textbf{c}) shows the case $N_\mathrm{p}=N/2$. We
evaluate $E/N_\mathrm{p}$ by MCSs, whereas
$E_\mathrm{cond}/N_\mathrm{p}$ is given by Eq.~(\ref{MFsubqA}). Also
here, $M_\mathrm{cond}/M\to 0$ and a QPT takes place at
$g_\mathrm{c}=2$ in agreement with Eqs.~(\ref{QPT}-\ref{QCP}).  In
\ref{OBCPBC} we report the hard-core boson case with PBC.  These
models could also be analyzed by mapping via the Jordan-Wigner
transformations~\cite{LSM,SuzukiBook} to the 1D $XXZ$ Heisenberg model
which, in turn, can be exactly solved by Bethe Ansatz~\cite{Baxter}.
In fact, the GS of the case $N_\mathrm{p}=N/2$ corresponds to the GS
of the $XXZ$ model, which changes character at the isotropic
ferromagnetic point~\cite{Baxter,Affleck} corresponding to
$g_\mathrm{c}=2$. \rev{More precisely, in the case $N_\mathrm{p}=N/2$
  the model (\ref{FermionHA}) maps to the $XXZ$ model in the sector of
  null magnetization $M^z=0$.
  Clearly, the constraint $M^z=0$ implies that there is not
  the ordinary up-down symmetry breaking, however,
  Eqs.~(\ref{QPT}-\ref{QCP}) allow to easily look for a condensation
  consisting in the formation of a closest packed configuration of
  fermions. It is however worth to observe that another kind of
  symmetry breaking can occur as the GS of (\ref{FermionHA}) is
  $N-N_\mathrm{p}$ degenerate.}

\section{1D Ising Model as a counter-example}
Our theory detects only first-order QPTs, consistently, we have to
check that no contradiction emerges when applied to a system which is
known to undergo a second-order QPT. Let us consider the 1D Ising
model ($N_\mathrm{p}=N$) with a transverse field of unitary amplitude
and PBC:
\begin{align}
  \label{Ising}
  H=-\sum_{i=1}^N\sigma_i^x - g
  \sum_{i=1}^{N}\sigma_i^z\sigma_{i+1}^z.
\end{align}
Here, $V_\mathrm{min}=-gN$, $M_\mathrm{cond}=2$, and
$\epsilon_\mathrm{cond}=-g$.  On the other hand, the model is exactly
solvable~\cite{Pfeuty} and for $N\to\infty$
\begin{align}
  \label{Ising1}
  \epsilon(g)=-\frac{1}{2\pi}\int_{-\pi}^\pi dq
  \left[1+2g\cos(q)+g^2\right]^{\frac{1}{2}},
\end{align}
which has a singularity of the second order at $g=1$ (i.e.,
$\epsilon'(g)$ is continuous, while $\epsilon''(g)$ is singular at
$g=1$).  As apparent from Fig.~\ref{fig1}(\textbf{d}), see
\ref{second_order_QPT} for a more quantitative survey, while
Eq.~(\ref{QPT}) is satisfied, Eq.~(\ref{QCP}) has no solution for
finite $g$, as the system always remains in the normal phase:
$\epsilon=\epsilon_\mathrm{norm} < \epsilon_\mathrm{cond}$.  In other
words, when the QPT is second order, Eq.~(\ref{QPT}) realizes only
through the equality $\epsilon=\epsilon_\mathrm{norm}$ being
$\epsilon_\mathrm{norm} < \epsilon_\mathrm{cond}$ $~\forall g$.

\rev{
  \section{On the fidelity approach and Anderson's orthogonality
    catastrophe}
  Fidelity, i.e., the absolute value of the overlap between two GSs
  evaluated at two different values of the Hamiltonian parameters,
  $F(g,g')=|\langle E(g)|E(g')\rangle|$, can be used to analyze a
  broad spectra of QPTs, including first-order QPTs, as well as cases
  where, as in our theory, there is no a priori knowledge of the order
  parameter~\cite{Fidelity_Gu}. The Fidelity approach looks for the
  minimum of $F(g,g+\delta g)$ with $\delta g$ small and fixed, In
  fact, the main idea is that near the critical point $g_\mathrm{c}$,
  the overlap between the GSs at $g<g_\mathrm{c}$ and at
  $g+\delta g>g_\mathrm{c}$, is minimal, and possibly zero, because of
  the symmetries (in a broad sense of the term ``symmetry'')
  associated to the two GSs are different.  In this respect, our
  theory is perfectly compatible with the fidelity approach. In fact,
  since
  $\mathbb{F}=\mathbb{F}_\mathrm{norm}\oplus
  \mathbb{F}_\mathrm{cond}$, by construction we have
  $\langle E_\mathrm{norm}|E_\mathrm{cond}\rangle=0$ for any system
  size. On the other hand, Eq. (\ref{QPT}) tells us that, in the
  thermodynamic limit, the GS of the Hamiltonian is either
  $|E_\mathrm{norm}\rangle$ or $|E_\mathrm{cond}\rangle$, for
  $g<g_\mathrm{c}$ or $g>g_\mathrm{c}$, respectively, so that, if
  Eq. (\ref{QCP}) has a solution, we conclude that, in the
  thermodynamic limit, the fidelity at the critical point is
  zero. Quite interestingly, in our theory the orthogonality between
  the two GSs is guaranteed to be exactly realized, i.e, $F(g,g')=0$, for any pair $(g,g')$
  whenever $g<g_\mathrm{c}$ and $g'>g_\mathrm{c}$. As it has also been
  pointed out in Ref.~\cite{Fidelity_Gu}, this rigid many-body
  orthogonality that takes place in the thermodynamic limit, has a
  famous phenomenology known as Anderson's orthogonality
  catastrophe~\cite{Anderson}. It is also quite interesting to observe
  that, in the model originally considered by Anderson, the rigid
  orthogonality is reached by replacing one single atom of the lattice
  host by an impurity atom.  In other words, in the thermodynamic
  limit the orthogonality is attained via an infinite dilution of the
  impurity, which is in parallel with the condition
  $M_\mathrm{cond}/M\to 0$ at the base of our theory.

  When compared to the fidelity approach, our theory offers a narrower
  spectra of applications, as it only applies to first-order QPTs.
  However, in detecting these latter, our method is numerically much
  more efficient than the fidelity approach. In fact, in our theory we
  analyze the QPT via the knowledge of the GS energies, whereas for
  evaluating the fidelity one needs the GSs, which, computationally,
  represent a much more demanding target~\cite{Fidelity_Gu}.}

\section{Conclusions}
\label{sec::Conclusions}
In conclusion, we have tested and verified Eqs.~(\ref{QPT}-\ref{QCP})
on a variety of models where a first-order QPT takes place.  The
mechanism at the basis of these QPTs is explained in terms of an
effective splitting of the Hilbert space
$\mathbb{F}=\mathbb{F}_\mathrm{norm}\oplus \mathbb{F}_\mathrm{cond}$
triggered by the condition $M_\mathrm{cond}/M\to 0$, with a normal,
classically intuitive phase, where
$\epsilon=\epsilon_\mathrm{norm}<\epsilon_\mathrm{cond}$, the system
being spread over $\mathbb{F}_\mathrm{norm}$, and a many-body
condensed, counter-intuitive phase, where
$\epsilon=\epsilon_\mathrm{cond}\leq \epsilon_\mathrm{norm}$, the
system being confined in $\mathbb{F}_\mathrm{cond}$.

\rev{In fact, the GS energy $E$, as the smallest eigenvalue of the
  Hamiltonian matrix in the configurational basis
  $\{|n\rangle\}_{n=1}^M$ is, in general, highly sensitive to a change
  of the matrix elements $\langle n' |H| n\rangle$. It is therefore
  intuitive to expect that, only the restriction of this matrix to a subset
  of configurations $\mathbb{F}_\mathrm{norm}$ that differ from
  $\mathbb{F}$ for an infinitesimal relative number of
  configurations can provide a smallest eigenvalue $E_\mathrm{norm}$
  in good approximation to $E$.  In the thermodynamic limit
  this classical guess translates as
  $\epsilon=\epsilon_\mathrm{norm}$.  However, such a naive intuition
  turns out to be wrong, in general. In the thermodynamic limit, the
  restriction of $H$ to an infinitesimal portion of the space, namely,
  $\mathbb{F}_\mathrm{cond}$, can actually determine completely the GS
  in a whole region of the Hamiltonian parameters and provide
  $\epsilon=\epsilon_\mathrm{cond} < \epsilon_\mathrm{norm}$.  This is
  a quite counter-intuitive behavior as in the case of the
  many-body Anderson's orthogonality catastrophe.  }

In the models
considered here, $\epsilon_\mathrm{cond}$ is found analytically,
whereas $\epsilon_{}$ or $\epsilon_\mathrm{norm}$ are evaluated by NDs
or MCSs.  In any case, $\epsilon_\mathrm{norm}$ and
$\epsilon_\mathrm{cond}$ are defined as GS energies of the Hamiltonian
$H$ of the system in the subspaces $\mathbb{F}_\mathrm{norm}$ and
$\mathbb{F}_\mathrm{cond}$, and, as such, represent a much easier
target than finding the first excited level of $H$ in the whole space
$\mathbb{F}$.  The class of QPTs that can be understood in terms of
first-order condensations via Eqs.~(\ref{QPT}-\ref{QCP}) is vast and
the method used here efficient.  We envisage several generalizations
and applications. In particular, the space $\mathbb{F}_\mathrm{cond}$
can be extended to include states corresponding to several low-energy
eigenvalues of $V$, not only the lowest one, as considered in the
present paper.  In this way, one can study systems with repulsive
long-range interactions and discover that phenomena like the so called
Wigner crystallization are in fact phase transitions belonging to the
present class of QPTs~\cite{WC}.

\ack M. O. acknowledges Capes PNPD and Grant CNPq 09/2018 - PQ
(Brazil).  We thank T. Macr\`{i} and J. Vitti for useful discussions.
We acknowledge also the cluster facility ``Jarvis'' at the Dep. of
Physics of UFSC - Brazil.

\appendix

\section{Specularity of  Eq.~(\ref{QPT}) and counter-examples}
\label{Spec}
According to Eq.~(\ref{QPT}), if, in the thermodynamic limit,
$M_\mathrm{cond}/M\to 0$, we have a sufficient condition to conclude
that $\epsilon=\min\{\epsilon_\mathrm{norm},\epsilon_\mathrm{cond}\}$.
However, even if $M_\mathrm{cond}/M \not\to 0$, it may still happen
that $\epsilon$ is the minimum of two quantities.  In fact, on
switching the roles of the operators $K$ and $V$ in $H$ (for
simplicity of notation, the parameter $g$ is now included in the
definition of $V$), i.e., writing $H=K'+V'$, with $K'=V$ and $V'=K$,
if $M'_\mathrm{cond}/M\to 0$, where $M'_\mathrm{cond}$ is the
dimension of the subspace where $V'$ is minimum, we still have
$\epsilon=\min\{\epsilon'_\mathrm{norm},\epsilon'_\mathrm{cond}\}$.
Let us consider three illustrative examples of this specularity
phenomenon.

\subsection{Modified Grover Model - Specularity with no QPT}
\label{Spec1}
\begin{figure}[t]
  \centerline{
  \includegraphics[width=0.7\columnwidth,clip]{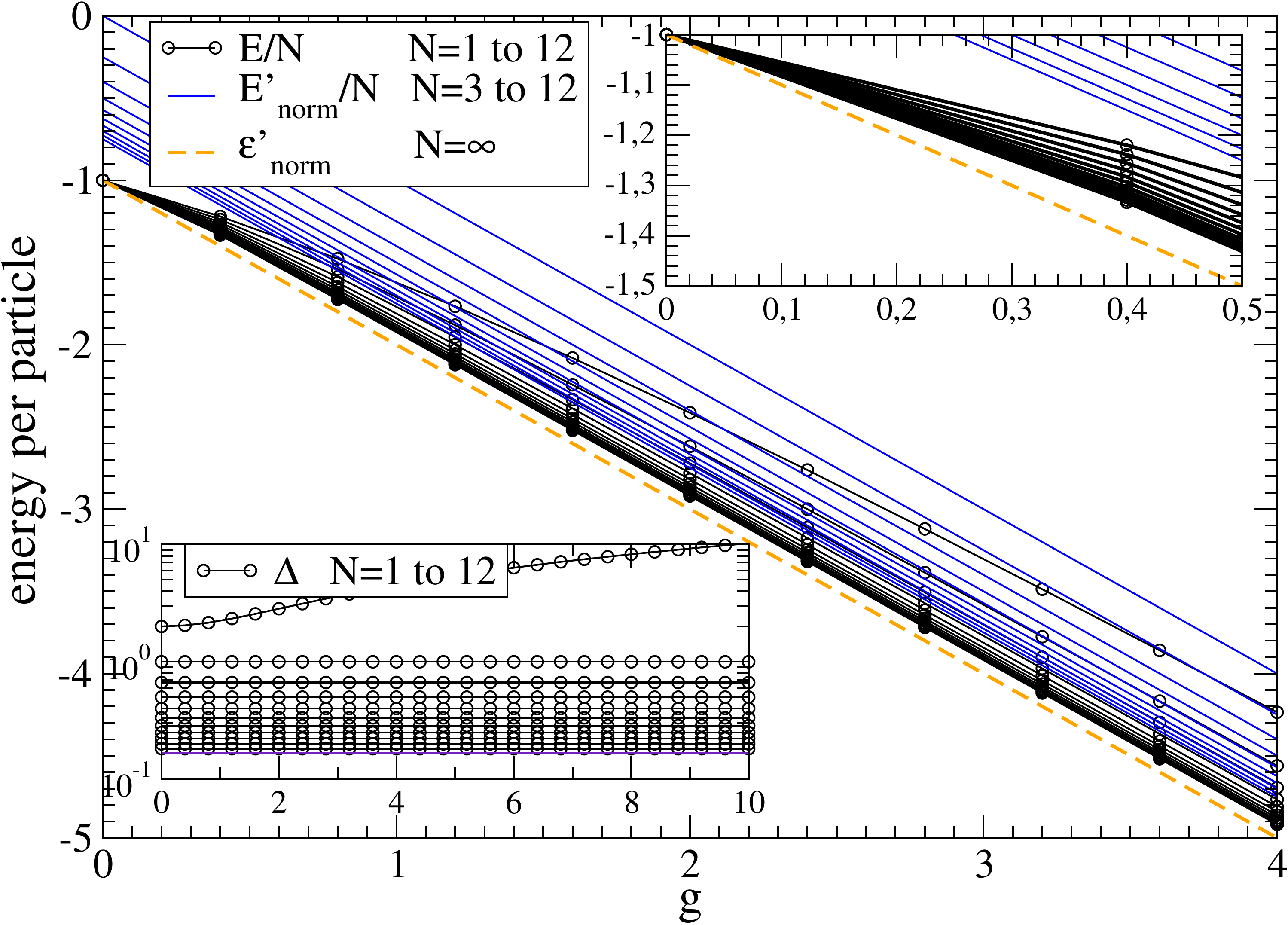}}
  \caption{GS energies per particle as a function of $g$ for the model
    described by Eq.~(\ref{GroverMod}), ``Modified Grover'' ($N_p=N$).
    Here, $E'_\mathrm{norm}/N=-1-g+3/N$ and $E/N$ is obtained by exact
    diagonalization. Upper Inset: particular of the plot in the range
    $g\in[0,0.5]$.  Lower Inset: gap $\Delta$ as a function of ${g}$,
    for $N=1$ to $N=12$ (highest to lowest plot).}
  \label{figa1}
\end{figure}

Let us introduce a modified version of the Grover Model as follows
(here, $N_\mathrm{p}=N$ and $M=2^N$)
\begin{align}
  \label{GroverMod}
  H=T_{N}-gN\frac{1+\sigma^z_1}{2}\bigotimes_{i=2}^N I_i,
  \qquad T_N=-\sum_{i=1}^{N}\sigma_i^x.
\end{align}
Let us indicate with $\ket{E_{T_N}^{(k)}}$, for $k=0,\ldots,N$ a
generic eigenstate of $T_N$ with eigenvalue $E_{T_N}^{(k)}=-(N-2k)$
(the degeneracy of the levels for $k\neq 0,N$ is not relevant for our
discussion).  Let us also indicate by $\ket{\uparrow^h}$ and
$\ket{\downarrow^h}$ the eigenstates of $\sigma^h$ with eigenvalues
$+1$ and $-1$, respectively, for $h=x,y,z$.  If, from
Eq.~(\ref{GroverMod}), we identify $K=T_{N}$, and
$V=-gN\ket{\uparrow^z}\bra{\uparrow^z}\otimes_{i=2}^N I_i$, we
have $V_\mathrm{min}=-gN$ and
$\mathbb{F}_\mathrm{cond}=\Span\{\ket{\uparrow^z}\otimes\ket{u}\}$,
where $\ket{u}$ is an arbitrary state of $N-1$ spins.  Therefore, in
this case we have $M_\mathrm{cond}=2^{N-1}=M/2$, 
  $\mathbb{F}_\mathrm{norm}=\Span\{\ket{\downarrow^z}\otimes\ket{v}\}$,
  with $\ket{v}$ an arbitrary state of $N-1$ spins and
\begin{align}
  \label{GroverMod1}
  &\ket{E_\mathrm{cond}}=\ket{\uparrow^z}\otimes\ket{E_{T_{N-1}}^{(0)}},
  \quad E_\mathrm{cond}/N=-1-g+\frac{1}{N}, \\
  &\ket{E_\mathrm{norm}}=\ket{\downarrow^z}\otimes\ket{E_{T_{N-1}}^{(0)}},
  \quad E_\mathrm{norm}/N=-1+\frac{1}{N} .
\end{align}
Hence, $\min\{\epsilon_\mathrm{norm},\epsilon_\mathrm{cond}\}=-1-g$.
However, we cannot conclude that $\epsilon=-1-g$ since
$M_\mathrm{cond}/M\to 1/2$ and the condition of Eq.~(\ref{QPT}) does not
apply.  On the other hand, if we exchange the role between $K$ and $V$
and choose $H=K'+V'$, with $K'=-gN\ket{\uparrow^z}
\bra{\uparrow^z}\otimes_{i=2}^N I_i$ and $V'=T_N$, we have
$V'_\mathrm{min}=-N$ and
$\mathbb{F}'_\mathrm{cond}=\Span\{\ket{E_{T_{N}}^{(0)}}\} $.
Therefore, in this case we have $M'_\mathrm{cond}=1$, so that Eq.~(\ref{QPT})
is valid and
$\epsilon=\min\{\epsilon'_\mathrm{norm},\epsilon'_\mathrm{cond}\}$.
Let us calculate the energies $\epsilon'_\mathrm{norm}$ and
$\epsilon'_\mathrm{cond}$. Since $\ket{E_{T_{N}}^{(0)}} =
\ket{\uparrow^x}\otimes\dots\otimes\ket{\uparrow^x}$, we have
\begin{align}
  \label{GroverMod2}
  &\ket{E'_\mathrm{cond}}=\ket{E_{T_{N}}^{(0)}}, \quad
    E'_\mathrm{cond}/N=-1-\frac{g}{2} , \\
  &\ket{E'_\mathrm{norm}}=
  \ket{\uparrow^z} \otimes \ket{E_{T_{N-1}}^{(1)}} , \quad
  E'_\mathrm{norm}/N=-1-g+\frac{3}{N} .
\end{align}
We conclude that
$\epsilon=\min\{\epsilon_\mathrm{norm}',\epsilon_\mathrm{cond}'\}
=-1-g$.  We have thus reached the same value for
$\min\{\epsilon_\mathrm{norm},\epsilon_\mathrm{cond}\}$ and
$\min\{\epsilon'_\mathrm{norm},\epsilon'_\mathrm{cond}\}$, however, in
the latter case we are able to identify this value with
$\epsilon$. Clearly, in the present model, by varying $g$ we find that
$\epsilon'_\mathrm{norm}$ is always smaller than
$\epsilon'_\mathrm{cond}$ and no QPT takes place, see Fig.~\ref{figa1}.

\subsection{Fermions in a heterogeneous external field - Specularity
  with QPT}
\label{Spec2}
\begin{figure}
  \centerline{
  \includegraphics[width=0.7\columnwidth,clip]{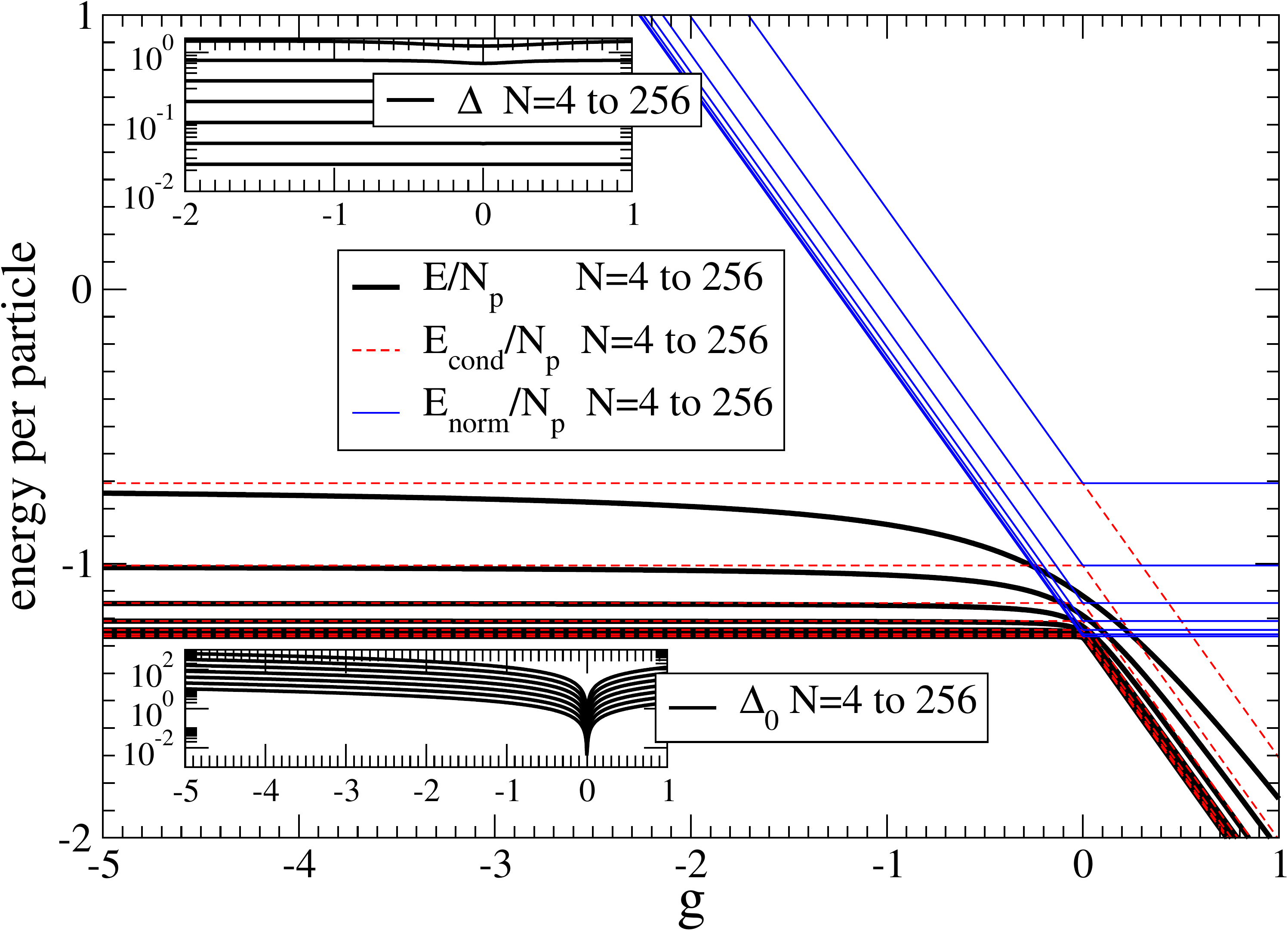}}
  \caption{GS energies per particle as a function of ${g}$ for the
    model described by Eq. (\ref{FermionHMod}) (spinless fermions with
    an extensive non uniform external field) at half-filling.  Here
    $E_\mathrm{cond}$ is given by Eq.(\ref{eFsubMod}) and $E$ is
    obtained by exact diagonalization.  Upper Inset: gap $\Delta(N)$
    for $N=4$ to $N=256$ (highest to lowest plot).  Lower Inset:
    $\Delta_1$ as a function of ${g}$ for $N=4$ to $N=256$ (highest to
    lowest plot).  It turns out that, at half-filling, the common
    discontinuity of $E_\mathrm{cond}$ and $E_\mathrm{norm}$ at $g=0$
    gets exactly canceled (which explains why in the present case we
    have continuous plots). }
  \label{figa2}
\end{figure}

Let us consider $N_\mathrm{p}$ fermions in a 1D chain of $N\geq
N_\mathrm{p}$ sites with open boundary conditions (OBCs) governed by a
Hamiltonian which is a simple modification of Eq.~(\ref{FermionH}), namely,
\begin{align}
  \label{FermionHMod}
  H=-\sum_{i=1}^{N-1}\left(c^\dag_ic_{i+1}+c^\dag_{i+1}c_i\right)
  -{g}N_\mathrm{p}c^\dag_1c_1.
\end{align}
If $K$ and $V$ correspond to the first and second term of
Eq.~(\ref{FermionHMod}), respectively, we can analyze this model as
done above by setting $N_\mathrm{imp}=1$ and replacing $g$
with $gN_\mathrm{p}$.  In particular, from Eqs.~(12) and (13) it now
follows
\begin{align}
  \label{MFsubMod}
  M_\mathrm{cond}=\frac{N_\mathrm{p}}{N}M, \qquad
  M=\binom{N}{N_\mathrm{p}},
\end{align}
\begin{align}
  \label{eFsubMod}
  E_\mathrm{cond}/N_\mathrm{p} = \left\{
    \begin{array}{ll}
      E^{(0)}(N-1,N_\mathrm{p})/N_\mathrm{p}, &\quad  g<0
      \\
      -g+E^{(0)}(N-1,N_\mathrm{p}-1)/N_\mathrm{p} , &\quad g>0
    \end{array}
  \right.
\end{align}
\begin{align}
  E_\mathrm{norm}/N_\mathrm{p} = \left\{
    \begin{array}{ll}
      -g+E^{(0)}(N-1,N_\mathrm{p}-1)/N_\mathrm{p}, &\quad g<0,
      \\
      E^{(0)}(N-1,N_\mathrm{p})/N_\mathrm{p}, &\quad g>0. 
    \end{array}
  \right.
\end{align}
In Fig.~\ref{figa2} we show the analysis of this model for $g\in[-2,1]$
in the half-filling case $N_\mathrm{p}=N/2$.  Despite the fact that
$M_\mathrm{cond}=M/2$, we have
$\epsilon=\min\{\epsilon_\mathrm{cond},\epsilon_\mathrm{norm}\}$. As
in the previous case, this is explained by switching the role between
$K$ and $V$, and observing that $M'_\mathrm{cond}=1$.  Note that now
we have a first-order QPT that takes place at $g_\mathrm{c}=0$. Quite
interestingly, in this QPT the gap $\Delta$ does not take any minimum
in correspondence of the critical point, and, for given $N$, remains
constant (see the discussion in the Introduction).

\subsection{Counter Example}
\label{Spec3}
From the previous examples, it turns out to be clear that, if we want
to find a case where $\epsilon <
\min\{\epsilon_\mathrm{norm},\epsilon_\mathrm{cond}\}$, as well as
$\epsilon < \min\{\epsilon'_\mathrm{norm},\epsilon'_\mathrm{cond}\}$,
we have to control that both $M_\mathrm{cond}/M \not\to 0$ and
$M_\mathrm{cond}'/M \not\to 0$.  A very simple model where this
occurs, is a system of $N$ spins in which only one of them is not free
and is subject to an extensive external field and an extensive
hopping:
\begin{align}
  \label{CE}
  H=-N \ket{\uparrow^x} \bra{\uparrow^x}\bigotimes_{i=2}^N I_i
  -gN \ket{\uparrow^z}\bra{\uparrow^z}\bigotimes_{i=2}^N I_i.
\end{align}
\rev{If} we identify as $K$ and $V$ the first and second terms in
Eq.~(\ref{CE}), respectively, we have $V_\mathrm{min}=-gN$ and
$\mathbb{F}_\mathrm{cond}=\Span\{\ket{\uparrow^z}\otimes\ket{u}\}$,
where $\ket{u}$ is an arbitrary state of $N-1$ spins.  Therefore, in
this case we have $M_\mathrm{cond}=M/2$ and
\begin{align}
  \label{CE1a}
  &\ket{E_\mathrm{cond}}=\ket{\uparrow^z}\otimes\ket{u}, \quad
    E_\mathrm{cond}/N=-\frac{1}{2}-g, \\
  \label{CE1b}
  &\ket{E_\mathrm{norm}}=\ket{\downarrow^z}\otimes\ket{v}, \quad
  E_\mathrm{norm}/N=-\frac{1}{2},
\end{align}
where $\ket{u}$ and $\ket{v}$ are two arbitrary states of $N-1$ spins.
On the other hand, if we define $K'=V$ and $V'=K$, we have
$V'_\mathrm{min}=-N$ and
$\mathbb{F}'_\mathrm{cond}=\Span\{\ket{\uparrow^x}\otimes\ket{u}\}$,
where $\ket{u}$ is an arbitrary state of $N-1$ spins.  It follows that
$M'_\mathrm{cond}=M/2$ and
\begin{align}
  \label{CE2a}
  &\ket{E_\mathrm{cond}'}=\ket{\uparrow^x}\otimes\ket{u}, \quad
    E'_\mathrm{cond}/N=-1-\frac{1}{2}g, \\
  \label{CE2b}
  &\ket{E'_\mathrm{norm}}=\ket{\downarrow^x}\otimes\ket{v}, \quad
  E'_\mathrm{norm}/N=-\frac{1}{2}g,
\end{align}
\rev{$\ket{v}$ being an arbitrary state of $N-1$ spins}.
Finally, we observe that the exact eigenvalues $E_{\pm}$ of the
Hamiltonian~(\ref{CE}) are easily calculated, the corresponding values
per particle being
\begin{align}
  \label{CE3}
  E_{\pm}/N=-\frac{1}{2}(1+g) \pm \frac{1}{2}\sqrt{1+g^2}.
\end{align}
We conclude that, as expected, the ground state energy per particle $E_-/N$ is, for any
value of $g>0$, strictly smaller than any of the energies given in
Eqs.~(\ref{CE1a})-(\ref{CE2b}), i.e., in the thermodynamic limit,
$\epsilon < \min\{\epsilon_\mathrm{norm},\allowbreak
\epsilon_\mathrm{cond},\allowbreak \epsilon'_\mathrm{norm},\allowbreak
\epsilon'_\mathrm{cond}\}$.

\subsection{Final remark}
It would be interesting to analyze more intermediate situations in
which $\min\{M_\mathrm{cond}/M, M'_\mathrm{cond}/M\}$ goes to zero
slowly in the thermodynamic limit, and to analyze how fast the error
obtained by assuming
$E/N_\mathrm{p}=\min\{E_\mathrm{cond}/N_\mathrm{p},\allowbreak
E_\mathrm{norm}/N_\mathrm{p},\allowbreak
E'_\mathrm{cond}/N_\mathrm{p},\allowbreak
E'_\mathrm{norm}/N_\mathrm{p}\}$ goes to zero in such limit.  This
will be the subject of future works.

\section{Comparing OBC with PBC}
\label{OBCPBC}
\begin{figure}[h]
  \centerline{
    \includegraphics[width=0.7\columnwidth,clip]{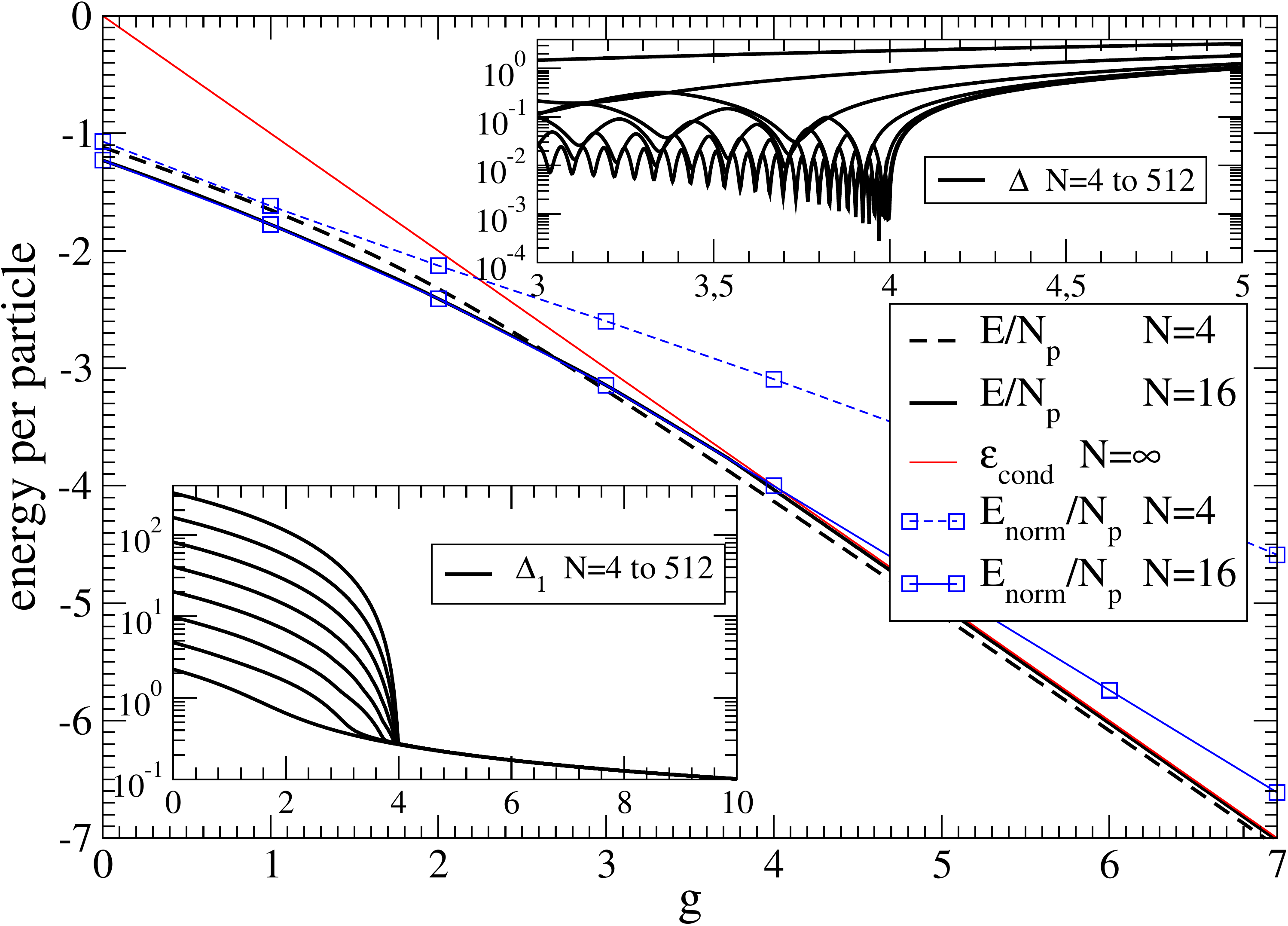}}
  \caption{GS energies per particle as a function of $g$ for the model
    described by Eq.~(\ref{FermionH}) (spinless fermions) with
    $N_\mathrm{p}=N_\mathrm{imp}=N/2$ in the case of OBC.  Here,
    $\epsilon_\mathrm{cond}=-g$ and $E$ is obtained by exact
    diagonalization.  Upper Inset: gap $\Delta$ as a function of $g$
    around the critical point $g_c\simeq 4.0$, for $N=4$ to $N=512$
    via powers of 2 (highest to lowest plot).  Lowest Inset: the
    function $\Delta_1$, for $N=4$ to $N=512$ via powers of 2 (lowest
    to highest).}
  \label{figa3}
\end{figure}

\begin{figure}[h]
  \centerline{
    \includegraphics[width=0.7\columnwidth,clip]{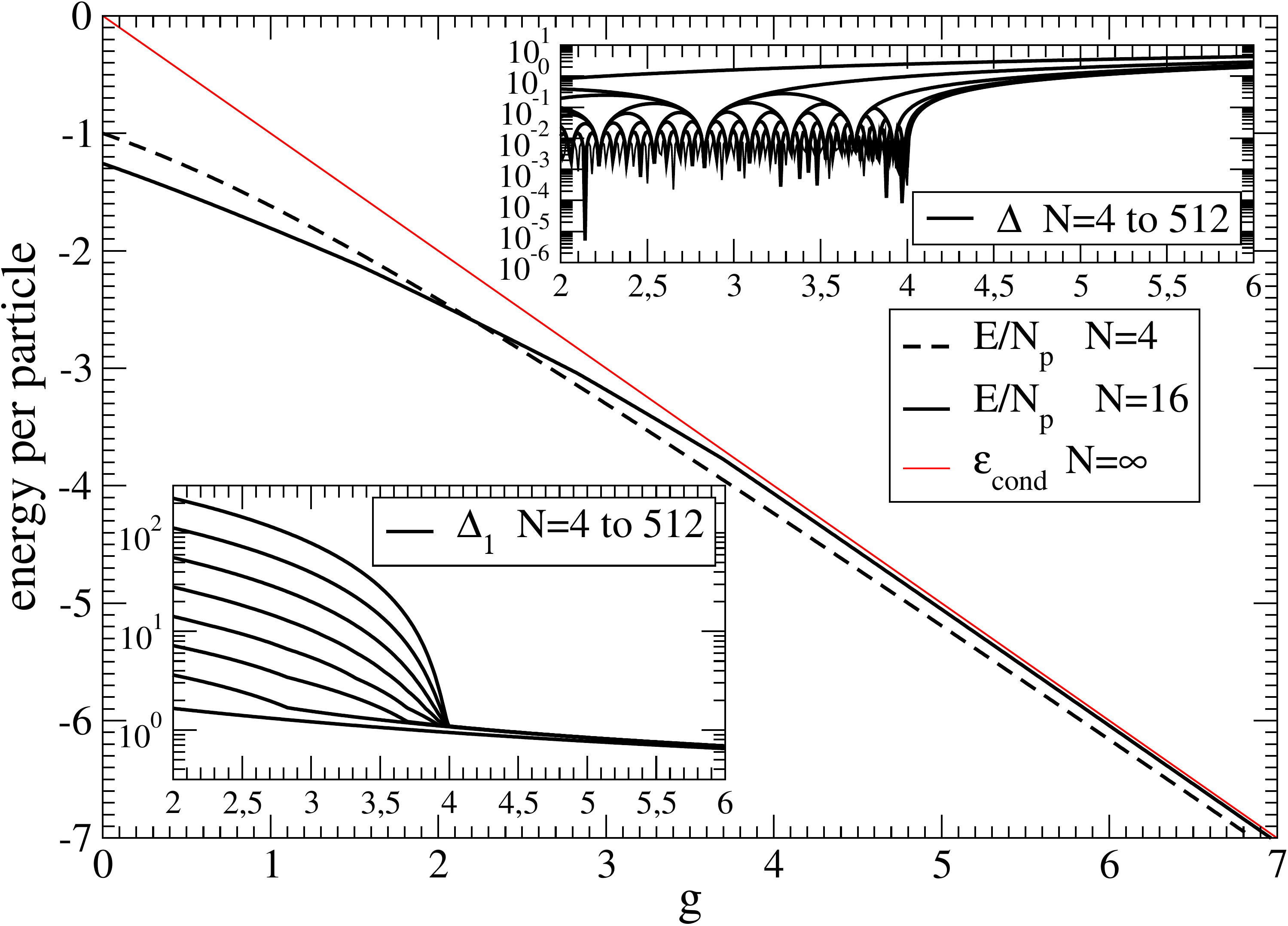}}
  \caption{As in Fig.~\ref{figa3} for PBC.}
  \label{figa4}
\end{figure}

Here, we elaborate on the model of Eq.~(\ref{FermionH}) and compare
the case with OBC, Fig.~\ref{figa3}, with the case with PBC,
Fig.~\ref{figa4}. We observe that only marginal differences emerge and
the critical point remains located in the same position of the OBC
case, $g_\mathrm{c}\simeq 4$.  In Section
\ref{spinless.fermions.with.external.field} we show a case with OBC to
avoid the sign problem which affects the MCS of any fermionic system,
except those in 1D with OBC.  In our case, this would affect the MCS
of $E_\mathrm{norm}$ (whereas $E$ and $E_\mathrm{cond}$ are evaluated
via exact diagonalization and analytically) reported in support of the
general theory, even tough, we actually locate the critical point by
means of $\Delta_1$, which does not make use of $E_\mathrm{norm}$.
Interestingly, as mentioned in Section
\ref{spinless.fermions.with.external.field}, we observe that the gap
$\Delta$ does not present a minimum in correspondence of
$g_\mathrm{c}$. In fact, for $g\to 0$, the GS energy of the model with
PBC becomes degenerate, causing a null gap in such a limit.  However,
as in all the other cases, $\Delta$ changes dramatically its character
when passing from the normal phase, $g<g_\mathrm{c}$, where it has a
wildly oscillating behavior, to the condensed phase, $g>g_\mathrm{c}$,
where it has a clear smooth behavior.

\begin{figure}[h]
  \centerline{
    \includegraphics[width=0.6\columnwidth,clip]{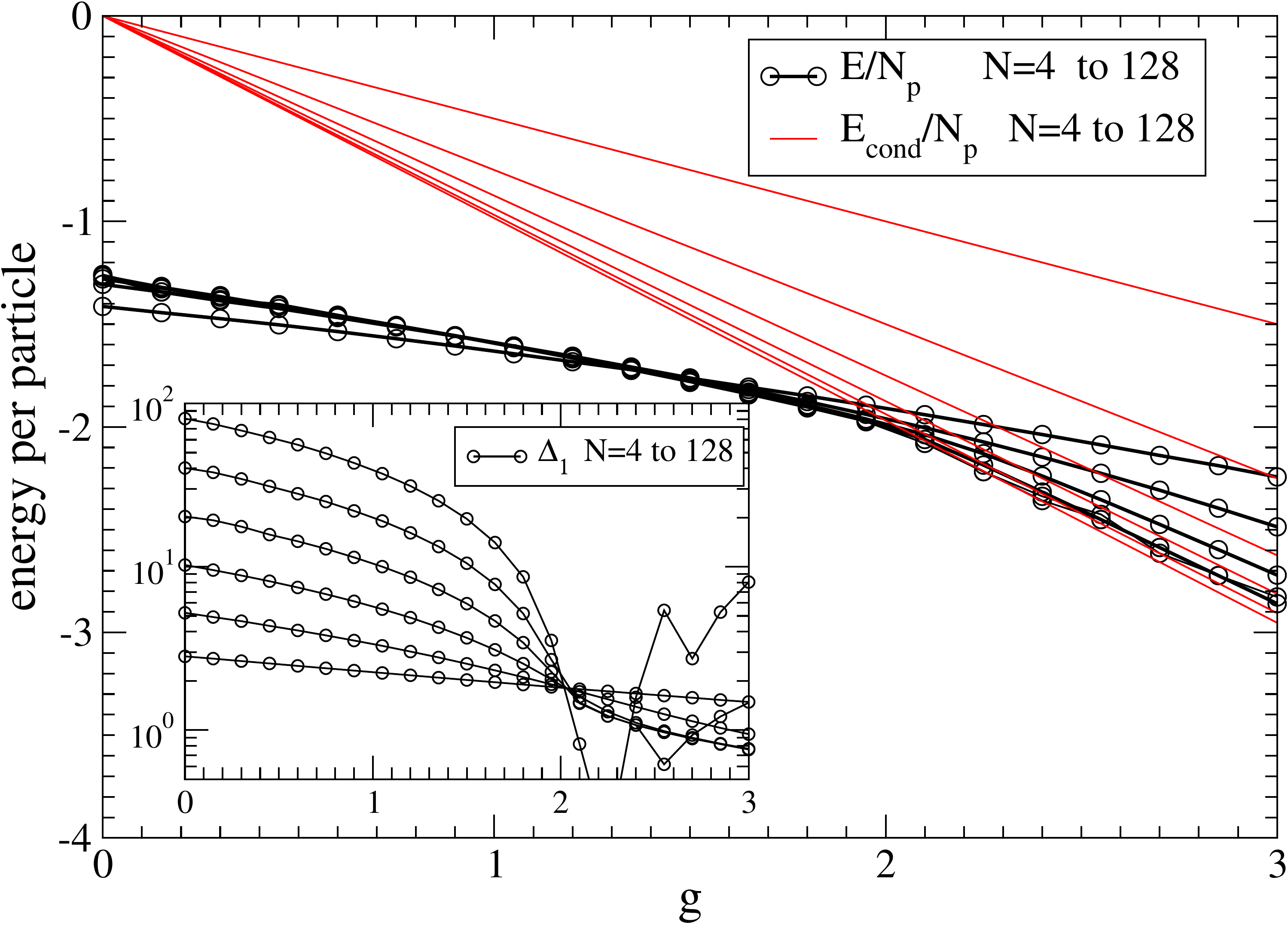}}
  \caption{GS energies per particle as a function of $g$ for the model
    described by Eq.~(\ref{FermionHA}) (hard-core bosons) with
    $N_\mathrm{p}=N/2$ in the case of PBC.  Here,
    $E_\mathrm{cond}/N_\mathrm{p}=-g(N_\mathrm{p}-1)/N_\mathrm{p}$ and
    $E$ is obtained by MCSs. Note that, at $g=0$, the lowest plot
    corresponds to the case $N=4$.  Inset: the function $\Delta_1$,
    for $N=4$ to $N=128$ via powers of 2 (lowest to highest). As in
    the case of Fig.~\ref{fig1}\textbf{c}, we see also here that the
    relation $E(N)/N_\mathrm{p} \leq E_\mathrm{cond}(N)/N_\mathrm{p}$
    is often violated for large $N$ and $g>g_\mathrm{c}$ due to the
    large fluctuations occurring in the MCSs (see discussion in
    Sec. \ref{MonteCarlo}).}
  \label{figa5}
\end{figure}

Similar considerations hold in the case of spinless fermions with an
attractive interaction, Eq.~(\ref{FermionHA}), compare
Fig.~\ref{fig1}\textbf{c} of Section
\ref{spinless.fermions.with.attractive.potential} with
Fig.~\ref{figa5}.

\section{A further non trivial example with a second-order QPT}
\label{second_order_QPT}
We have pointed out that our theory encoded in Eqs.~(\ref{QPT}) and
(\ref{QCP}), detects only first-order QPTs.  At the same time, we have
stressed that our Eq.~(\ref{QPT}), provided that
$M_\mathrm{cond}/M\to 0$, is an identity that holds in any situation,
regardless of any possible QPT and, in particular, regardless of the
existence of a solution of Eq.~(\ref{QCP}).  This fact has been made
concrete by showing the analysis of the 1D Ising model in the presence
of a transverse field, Eq.~(16) and Fig.~\ref{fig1}(\textbf{d}). To
further support our claim, we now consider a generalization of the 1D
Ising model that, besides the usual two-spin interaction, includes
also a four-spin interaction as follows
\begin{align}
  \label{Ising4}
  H=-\sum_{i=1}^N\sigma_i^x - g\sum_{i=1}^{N}\sigma_i^z\sigma_{i+1}^z
  -g'\sum_{i=1}^{N}\sigma_i^z\sigma_{i+1}^z\sigma_{i+2}^z\sigma_{i+3}^z,
  \qquad N\geq 4,
\end{align}
where $g'\geq 0$ is, besides $g$, a second free dimensionless
parameter and PBC are understood.  The aim of the present Appendix is
threefold: by making use of extensive NDs, we demonstrate that: (i)
Eq.~(\ref{QPT}) is satisfied, (ii) Eq.~(\ref{QCP}) has no finite
solution, and (iii) the system undergoes a second-order QPT.

\begin{figure}[h]
  \centerline{
    \includegraphics[width=0.7\columnwidth,clip]{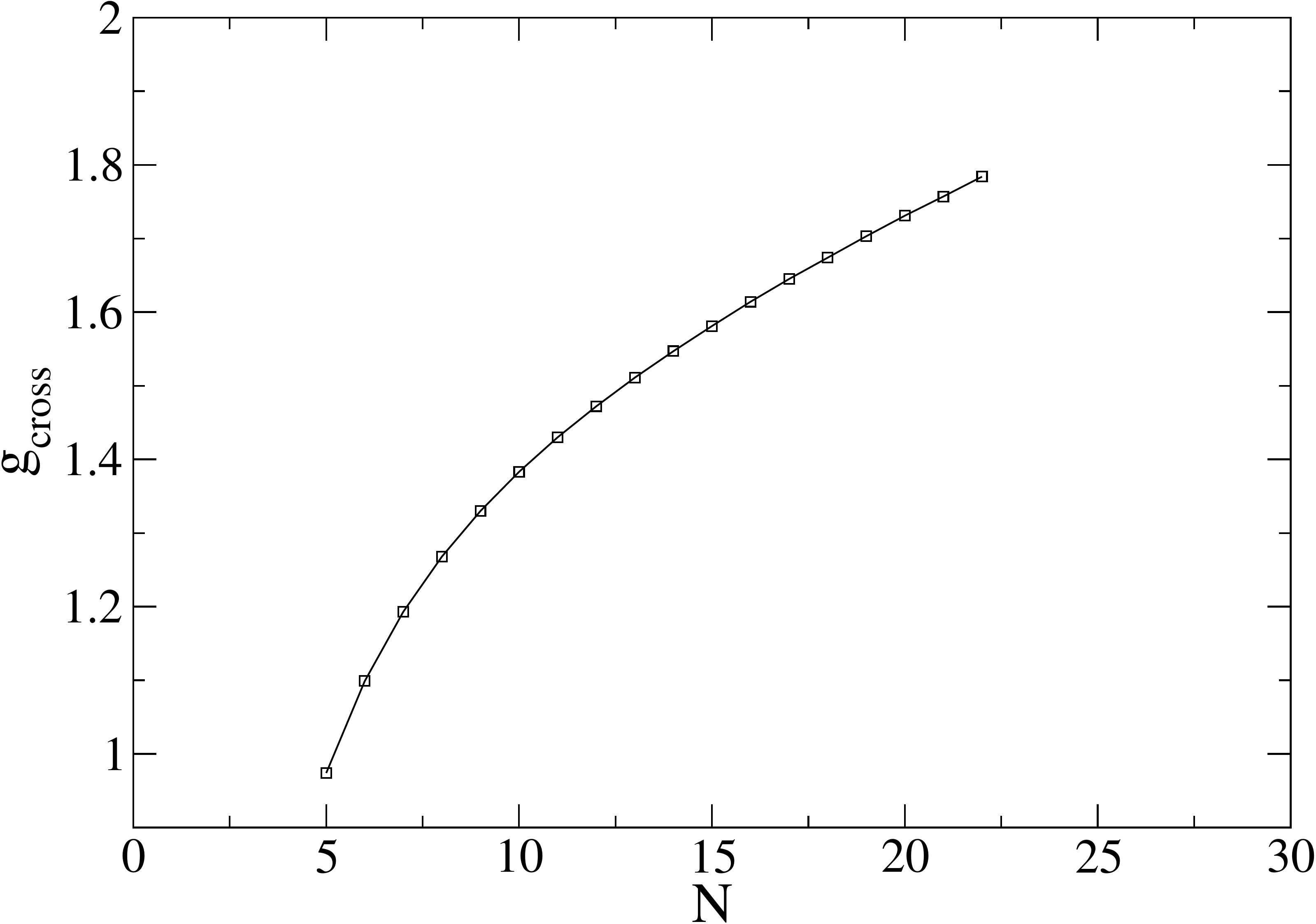}}
  \caption{Plot of $g_\mathrm{cross}(N)$ versus $N$ for the 1D Ising
    model (corresponding to the choice $g'=0$ in Eq. (\ref{Ising4}))
    for $N=5, 6, \ldots 22$. The crossing point $g_\mathrm{cross}(N)$
    is defined as the value of $g$ at which
    $E_\mathrm{norm}(N)/N=\epsilon_\mathrm{cond}$, where
    $\epsilon_\mathrm{cond}=-g$ and $E_\mathrm{norm}(N)$ is evaluated
    by NDs.}
  \label{figa6}
\end{figure}

\begin{figure}[h]
  \centerline{ \includegraphics[width=\columnwidth,clip]{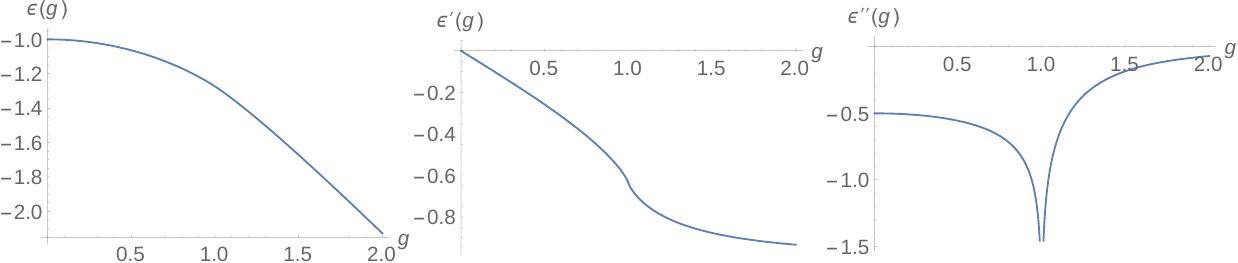}}
  \caption{From left to right: plots of $\epsilon(g)$ ,
    $\epsilon'(g)$, and $\epsilon''(g)$ versus $g$, where
    $\epsilon(g)$ is the exact GS energy per particle of the 1D Ising
    model ($g'=0$) in the thermodynamic limit, as given by Eq.~(\ref{Ising1}).}
  \label{figa7}
\end{figure}

\textit{Case $g'=0$}.  Before providing such demonstrations for the
general Hamiltonian~(\ref{Ising4}), it is useful to consider again the
Ising case $g'=0$.  From Fig.~\ref{fig1}(\textbf{d}) it is evident
that, for any $g$, $E_\mathrm{norm}(N)/N\to \epsilon$ for
$N\to\infty$.  Since $M_\mathrm{cond}/M\to 0$, one may wonder that the
result $\epsilon=\epsilon_\mathrm{norm}$ is a violation of
Eq.~(\ref{QPT}). However, this is not the case, Eq.~(\ref{QPT}) holds
true because we also have
$\epsilon_\mathrm{norm}< \epsilon_\mathrm{cond}$ for any $g$.  This
inequality can be numerically validated as follows.  At any finite
size $N$, the curves $E_\mathrm{norm}(N)/N$ and
$\epsilon_\mathrm{cond}$ cross, as a function of $g$, at the point
$g_\mathrm{cross}(N)$, see Fig.~\ref{fig1}(\textbf{d}).  From
Fig.~\ref{figa6}, where we plot $g_\mathrm{cross}(N)$ as a function of
$N$, it is evident that, after an initial transient,
$g_\mathrm{cross}(N)$ grows linearly with $N$.  The inequality
$\epsilon_\mathrm{norm}< \epsilon_\mathrm{cond}$ for any $g$, allows
also to conclude that Eq.~(\ref{QCP}) is not satisfied at any finite
$g$.  This excludes, therefore, the possibility of a first-order
QPT. It is well known, however, that the Ising model has a second-order
QPT at $g=1$. Whereas there exist several methods to show up the
second-order nature of this QPT, in our framework, where we mainly
work with the GS energy, it is convenient to analyze the nature of the
possible singularities of $\epsilon(g)$ with respect to $g$.  We
remind that a QPT transition is classified of first-order when the GS
energy per particle $\epsilon(g)$ has a jump discontinuity in its
first derivative $\epsilon'(g)=d\epsilon(g)/dg$, while it is
classified of second-order when
$\epsilon^{(k)}(g)=d^k \epsilon(g)/dg^k$ has a discontinuity for some
$k\geq 2$. Such a definition parallels the definition of classical,
finite temperature, QPT in terms of the thermodynamic limit of the
free energy per particle.  Figure~\ref{figa7} shows $\epsilon(g)$,
$\epsilon'(g)$ and $\epsilon''(g)$, where $\epsilon(g)$ is the exact
GS energy per particle of the 1D Ising model ($g'=0$) in the
thermodynamic limit, as given by Eq.~(\ref{Ising1}). Clearly, we are facing the
well known scenario of a second-order QPT, where $\epsilon'(g)$ is
continuous, while $\epsilon''(g)$ has a singularity at the critical
point $g_c=1$.

\begin{figure}[h]
  \centerline{
    \includegraphics[width=0.7\columnwidth,clip]{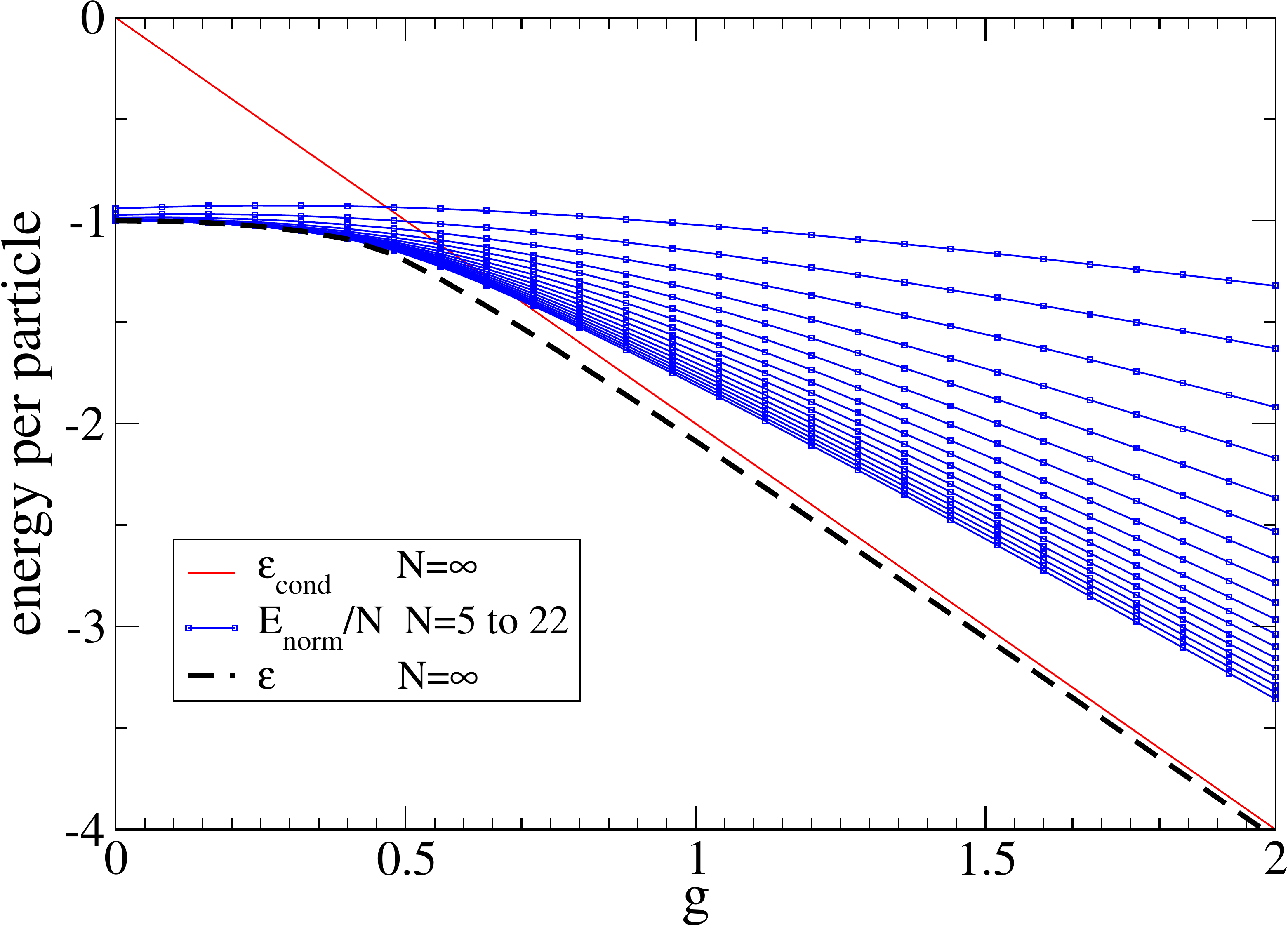}}
  \caption{GS energies per particle for the model with Hamiltonian
    (\ref{Ising4}) in the case $g'=g$: $\epsilon_\mathrm{cond}=-2g$
    (continuous line), $\epsilon$ (dashed line), $E_\mathrm{norm}/N$
    (line with points) for $N$= 5 to 22 (top to bottom). Here
    $\epsilon$ has been obtained by using $\epsilon\simeq E(N)/N$ for
    any $N$ large enough. In fact, as in the Ising case ($g'=0$), also
    in this case $E(N)/N\to \epsilon$ very quickly, the differences
    between $N=12$ and $N=13$, for example, being unrecognizable at
    the shown scale. Here, we have evaluated $\epsilon\simeq E(N)/N$
    with $N=22$.}
  \label{figa8}
\end{figure}

\begin{figure}[h]
  \centerline{
    \includegraphics[width=0.7\columnwidth,clip]{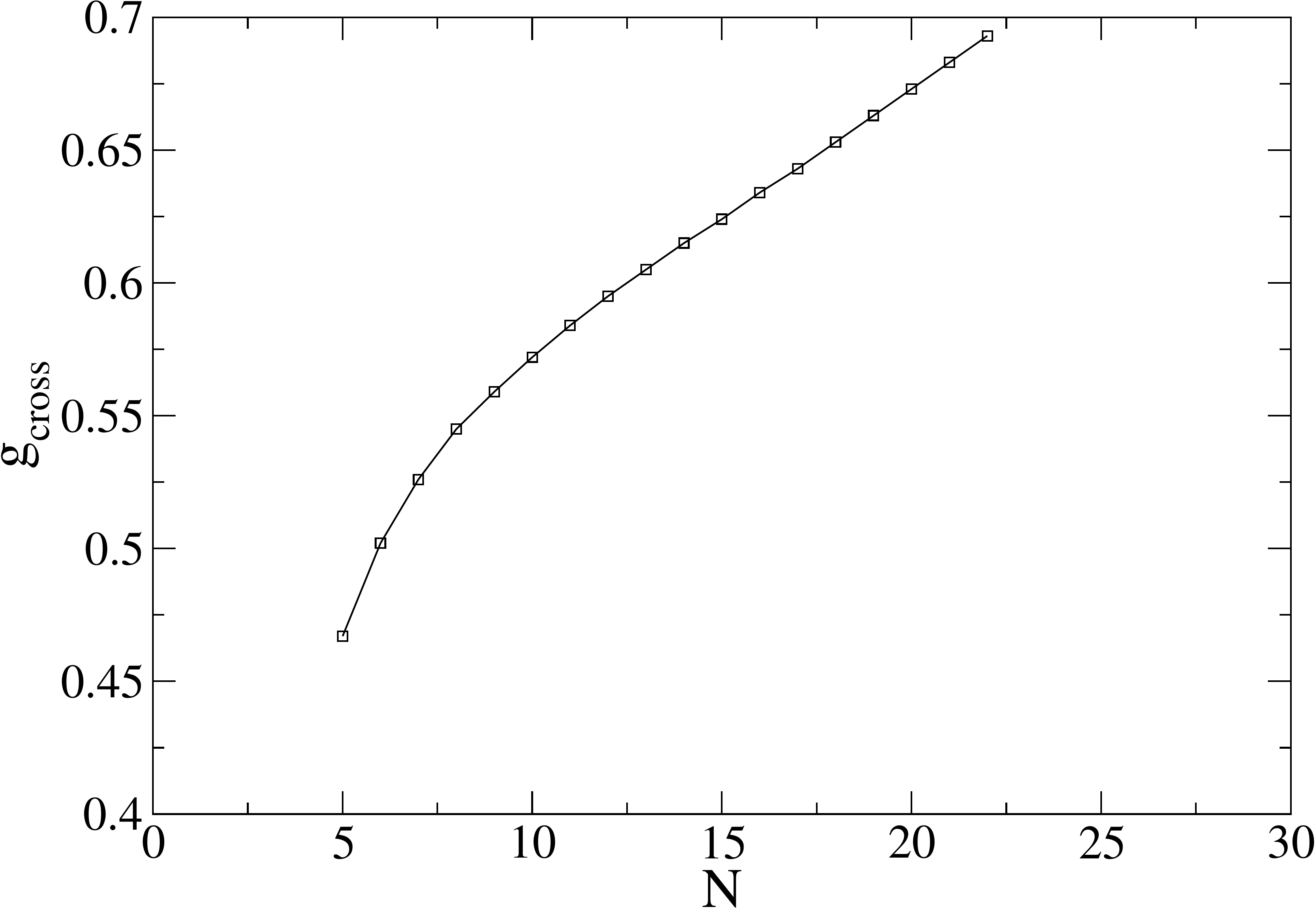}}
  \caption{Plot of $g_\mathrm{cross}(N)$ versus $N$ for the model with
    Hamiltonian (\ref{Ising4}) in the case $g'=g$ for
    $N=5, 6, \ldots 22$.  The crossing point $g_\mathrm{cross}(N)$ is
    defined as the value of $g$ at which
    $E_\mathrm{norm}(N)/N=\epsilon_\mathrm{cond}$, where
    $\epsilon_\mathrm{cond}=-2g$ and $E_\mathrm{norm}(N)$ is evaluated
    by NDs.}
  \label{figa9}
\end{figure}

\begin{figure}[h]
  \centerline{ \includegraphics[width=\columnwidth,clip]{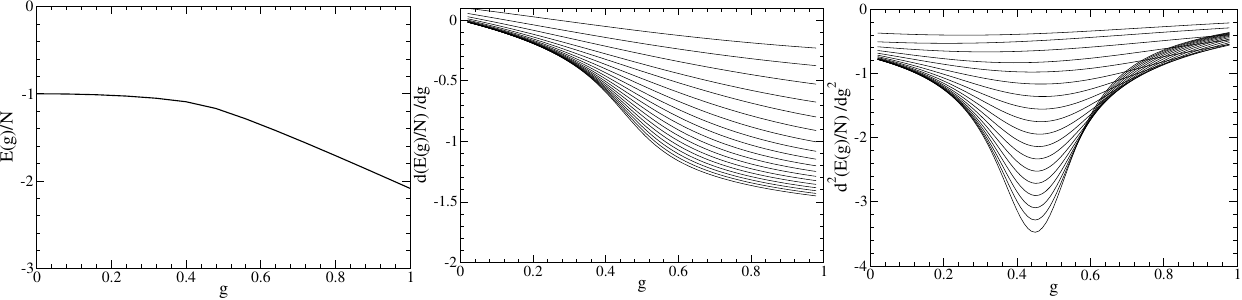}}
  \caption{ From left to right: plots of $E(g)/N$, $E'(g)/N$, and
    $E''(g)/N$ as a function of $g$ for $N=5$ to 22 for the model with
    Hamiltonian (\ref{Ising4}) obtained by NDs.  In the plot of
    $E(g)/N$ the curves obtained for different values of $N$ are
    indistinguishable.}
\label{fig8}
\end{figure}

\textit{Case $g'>0$}.  For $g'\neq 0$ the Hamiltonian~(\ref{Ising4})
is rather non trivial.  Nevertheless, in the ferromagnetic case
$g'>0$, the scenario obtained in the Ising case ($g'=0$) remains
essentially unchanged. For simplicity, we set $g'=g$.  The analysis
for different positive values of $g'$, not reported here, leads to the
same qualitative behavior.  In Figs.~\ref{figa8}, \ref{figa9}, and
\ref{fig8}, we see, respectively, the analogous of
Figs.~1(\textbf{d}), \ref{figa6}, and \ref{figa7} corresponding to the
Ising case $g'=0$.  It is evident that, also for $g'>0$, we have: (i)
Eq.~(\ref{QPT}) is satisfied, (ii) Eq.~(\ref{QCP}) has no finite
solution in terms of the parameter $g'=g$, and (iii) the system
undergoes a second-order QPT (in this case the critical point being
located near $g=0.44$).  It is worth to mention that the fact that the
observed QPT of this 1D model remains of second-order for any non
negative value of $g'$, is quite different from the mean-field case
where, at least classically, as is known, for suitable positive values
of $g$ and $g'$ one can have also first-order QPTs.

\section{Monte Carlo simulations}
\label{MonteCarlo}
The method used to perform our MCSs on lattice systems is based on an
exact probabilistic representation of the quantum dynamics via Poisson
processes that, virtually, reproduce the trajectories determined by
the hopping operator $K$~\cite{EPR,EPRMC}.  The corresponding Monte
Carlo sampling is exact in the sense that there are no systematic
errors due to any finite-time approximations (there is no Trotter
approximation, see, e.g., \cite{Ceperley}).  The GS energy of a system
governed by a Hamiltonian $H$ can then be obtained from the
evaluations of the matrix elements of the evolution operator
$\exp(-H t)$ at imaginary times $t$ in the limit $t\to +\infty$.  As
in any MCS, sampling the matrix elements of $\exp(-H t)$ involves
fluctuations that increase exponentially with $t$. These fluctuations
can be reduced by using a reconfiguration
technique~\cite{Hetherington,Sorella}: instead of following many
independent sample-trajectories that evolve during a long time $t$,
one follows the evolution of a set of $\mathcal{M}\gg 1$ simultaneous
trajectories that evolve along the shorter times $\Delta t=t/R$, where
$R$ is an integer sufficiently large to keep the fluctuation along
$\Delta t$ small. At the end of each time step $\Delta t$, the final
configurations with index $i=1,\dots,\mathcal{M}$ are given a suitable
weight $p_i$ which is used to generate randomly the initial
configurations of the subsequent time step. The procedure stops after
$R$ time steps.  In the limit $\mathcal{M}\to\infty$ this procedure
becomes exact (no bias is introduced) \cite{EPRMC}. By a suitable
choice of $\mathcal{M}$ and $R$ this technique allows us to handle the
MCS of our models even close to the critical points, where in
principle we should let $t\gg 1/\Delta$, where $\Delta$ is the gap of
the model.
\begin{table}
  \centering{
    \begin{tabular}{ l c c r }
      $N~$ & $N_\mathrm{p}~$ & $\Delta t~$ & $R~$
      \\ \hline
      4~ & 2~ & 16~ & 64~ \\ 
      8~ & 4~ & 16~ & 128~ \\ 
      16~ & 8~ & 16~ & 256~ \\ 
      32~ & 16~ & 32~ & 512~ \\ 
      64~ & 32~ & 32~ & 1024~ \\ 
    \end{tabular}}
  \caption{Statistical parameters used for the MCSs of 1D free
    fermions in a non-uniform external field
    (Fig.~\ref{fig1}\textbf{b} and Fig.~\ref{figa3}).}
\end{table}
\begin{table}
  \centering{
    \begin{tabular}{ l c c r }
      $N~$ & $N_\mathrm{p}~$ & $\Delta t~$ & $R~$ 
      \\\hline 
      4~ & 2~ & 16~ & 64~ \\
      8~ & 4~ & 16~ & 128~ \\
      16~ & 8~ & 16~ & 256~ \\
      32~ & 16~ & 64~ & 512~ \\
      64~ & 32~ & 64~ & 1024~ \\
      128~ & 64~ & 64~ & 2048~ \\
    \end{tabular}}
  \caption{Statistical parameters used for the MCSs of 1D interacting
    fermions and 1D hard-core bosons (Fig.~\ref{fig1}\textbf{c} and
    Fig.~\ref{figa5}).}
\end{table}

The above procedure cannot be applied for $\Delta t$ too small: below
a certain threshold of $\Delta t$, the system simply does not evolve.
In fact, given the hopping operator $K$, one must take into account
that the mean number of jumps $\media{N}_t$ of a virtual trajectory
along a time $t$ is, up to a dimensional factor that we set to 1 in
our models, $\media{N}_t= E^{(0)} t$, where $E^{(0)}$ is the GS energy
of the system without potential, i.e., the case with $g=0$.
Therefore, it is necessary to choose $R$ such that
$\Delta t E^{(0)}\geq 1$. In the absence of a QPT the optimal choice
corresponds to $\Delta t = 1/E^{(0)}\propto 1/N_\mathrm{p}$ which, in
the absence of any sign problem, allows to perform efficient
simulations for systems of large size~\cite{EPRMC}.  However, if the
model undergoes a QPT, such a choice works only far from the critical
point and larger values of $\Delta t$ must be considered.  Given the
magnitude of the desired maximal simulation times to be performed on
an ordinary PC, ranging in our cases from a few ours to a few days,
there is not a simple recipe to select the optimal values of
$\mathcal{M}$ and $R$, the best criterion being empirical with the
constrain $\Delta t E^{(0)}\geq 1$.  In Tables I and II we show the
statistical parameters chosen to perform our MCSs. In all cases we
have used a single set of $\mathcal{M}=2^{20}$ parallel
trajectories. Table I refers to Fig.~\ref{fig1}\textbf{b} and
Fig.~\ref{figa3}.  In these cases the MCSs have been used only for
evaluating $E_\mathrm{norm}(N)$, which actually is not used to locate
the critical point, but only to show (for a few system sizes $N$) how
the general theory takes effect.  Table II refers to the cases of
Fig.~\ref{fig1}\textbf{c} and Fig.~\ref{figa5}.  In all cases, as we
approach the region $g\geq g_\mathrm{c}$ the statistics becomes more
demanding, an issue which becomes more pronounced in the presence of
interaction (see the fluctuations in the Inset of
Fig.~\ref{fig1}\textbf{c} and in the Fig.~\ref{figa5} for
$g\geq g_\mathrm{c}$).  Indeed, a sign of the fact that, for
$g> g_\mathrm{c}$, the MCSs are affected by large fluctuations emerges
by observing that the relation $E(N)\leq E_\mathrm{cond}(N)$ is often
violated for large $N$ and $g>g_\mathrm{c}$.  However, this problem
does not prevent us to locate well the critical point $g_\mathrm{c}$
also in the presence of interaction. These large fluctuations could be
reduced by exploiting the partial information that we have about the
GS for $g>g_\mathrm{c}$ and using importance sampling, as explained
in~\cite{EPRMC}. Such a refinement is however beyond the aim of the
present work.

\section*{References}


\providecommand{\newblock}{}

\end{document}